\documentclass[twocolumn,prc,showpacs,preprintnumbers,
               unsortedaddress,amsmath,amssymb,floatfix]{revtex4}
\usepackage{graphicx}% Include figure files
\usepackage{dcolumn}% Align table columns on decimal point
\usepackage{bm}% bold math
\usepackage{longtable}

\newcommand{\beq}{\begin{equation}}
\newcommand{\eeq}{\end{equation}}
\newcommand{\beqn}{\begin{eqnarray}}
\newcommand{\eeqn}{\end{eqnarray}}
\newcommand{\btab}{\begin{tabular}}
\newcommand{\etab}{\end{tabular}}

\newcommand{\etal}{{\em{et al.}}}

\usepackage{multirow,amsmath,booktabs}

\begin{document}

\title{A New Equation of State for Astrophysical Simulations}

\author{G.~Shen\footnote{e-mail:  gshen@indiana.edu}}
\affiliation{Nuclear Theory Center and Department of Physics,
Indiana University Bloomington, IN 47405}
\author{C.~J.~Horowitz\footnote{e-mail:
horowit@indiana.edu} } \affiliation{Nuclear Theory Center and
Department of Physics, Indiana University Bloomington, IN 47405}
\author{S.~Teige\footnote{e-mail:
steige@indiana.edu}}
\affiliation{University Information Technology Services, Indiana University, Bloomington, IN 47408}
\date{\today}
\begin{abstract}
We generate a new complete equation of state (EOS) of nuclear
matter for a wide range of temperatures, densities, and proton
fractions ready for use in astrophysical simulations of supernovae and neutron star mergers. Our previous two papers tabulated the EOS at over 180,000 grid points in the temperature range $T$ = 0 to 80 MeV, the density range $n_B$ = 10$^{-8}$
to 1.6 fm$^{-3}$, and the proton fraction range $Y_P$ = 0
to 0.56. In this paper we combine these data points using a suitable interpolation scheme to generate a single equation of state table on a finer grid.  This table is thermodynamically consistent and conserves entropy during adiabatic compression tests. We present various thermodynamic quantities and the composition of matter in the new EOS, along with several comparisons with existing EOS tables.  Our equation of state table is available for download.  
\end{abstract}


\maketitle

\section{Introduction}

The equation of state (EOS), pressure as a function of density and temperature, for warm dense nuclear matter is an important ingredient in the theory of core collapse supernovae \cite{MPAreview}.  The EOS influences the shock formation and evolution, and determines the compactness of the nascent proto-neutron star.  The EOS also impacts the position of the neutrinosphere, which is the site of last scattering for neutrinos in supernovae, and the emitted neutrino spectra \cite{light07}. This helps determine neutrino flavor oscillations \cite{Duan10}, and ultimately the spectra of supernova neutrinos in terrestrial detectors.

In a previous paper \cite{SHT10a} we used a relativistic mean field (RMF) model to self-consistently calculate non-uniform matter at intermediate density and uniform matter at high density.  In a following paper \cite{SHT10b}, we used a Virial expansion for a nonideal gas of nucleons and nuclei to obtain the EOS at low densities.  Our Virial approach \cite{Virial1,Virial2,Virial3} uses elastic scattering phase shifts and nuclear masses as input.  We also include Coulomb corrections that can be important for neutrino interactions \cite{ioncorrelations}.  Alternative approaches to the equation of state at low densities, such as \cite{typel}, can include model dependent strong interactions.  Altogether our relativistic mean field and Virial EOS models cover the large range of temperatures, densities, and proton fractions shown in Table ~\ref{tab:phasespace}.

\begin{table}[h]
\centering \caption{Range of temperature $T$, baryon density $n_B$ and proton
fraction $Y_P$ in the EOS table.} \label{tab:phasespace}\btab{cccc}
\hline \hline
Parameter & minimum & maximum & number of points \\
 \hline
$T$ [MeV] & 0, 10$^{-0.8}$ & 10$^{1.9}$ & 36 \\

log$_{10}$(n$_B$) [fm$^{-3}$] &-8.0 & 0.2 & 83 \\

Y$_P$  & 0, 0.05  & 0.56 & 53 \\

\hline \etab
\end{table}

Note that pure neutron matter (Y$_P$=0) is also included in the table. The
equation of state at zero temperature is obtained by extrapolation
from the two lowest temperatures. There are 73,840 data points
from the Virial calculation at low densities, 17,021 data points from the nonuniform 
Hartree calculation, and 90,478 data points from uniform matter
calculations. The overall calculations took
7,000 CPU days in Indiana University's supercomputer clusters.

It is most efficient for astrophysical simulations of supernovae and neutron star mergers to interpolate within an existing EOS table to update the properties of stellar matter with three independent thermodynamics parameters ($T$, $n_B$ and $Y_P$). Therefore in this paper we generate, from these original results, a sufficiently large EOS table that can easily be interpolated in a way that preserves thermodynamic consistency.

There exist only two realistic EOS tables that are in widespread use for astrophysical simulations, the Lattimer-Swesty (L-S) equation of state \cite{LS}, that uses a compressible liquid drop model with a Skyrme force, and the H. Shen, Toki, Oyamatsu and Sumiyoshi (S-S) equation of state \cite{Shen98a,Shen98}, that uses the Thomas-Fermi and variational approximations with an RMF model.  The two EOS show clear differences for example in the nuclear composition during the core-collapse phase and in the adiabatic index around and above the phase transition to homogeneous nuclear matter \cite{MPAreview}. In this paper, we will discuss several comparisons between our new EOS and these two existing ones.

The paper is organized as follows: In section \ref{numerics} we
present the numerical details for the interpolation scheme we use to generate the full
EOS table.  Section \ref{result} shows results for our EOS, including various thermodynamic
properties and the composition. The $T$ = 0 beta
equilibrium EOS is also presented and used to calculate the neutron star
mass-radius relation.  We also give some comparisons between our EOS and the
two existing EOS tables.  Section \ref{summary} presents a summary of
our results and gives an outlook for future work.  In the appendix
\ref{app} the contents of our EOS table are explained in more detail and the
public access to our EOS table is given.

This paper presents results based on the relatively stiff RMF interaction of Lalazissis, K\"{o}nig, and Ring \cite{NL3} that has a high pressure at high densities.  In later work we will present additional EOS tables for other softer interactions.  This will allow one to more easily run astrophysical simulations for different EOS tables and identify quantities that are sensitive to the EOS.  Note that a variety of uncertainties in the EOS and some laboratory measurements and  astronomical observations related to the EOS have been discussed in our previous papers \cite{SHT10a,SHT10b}.

\section{\label{numerics}Numerical details}

In this section we describe how we interpolate our free energy results and calculate other quantities from derivatives of the free energy in a thermodynamically consistent manner.

\subsection{Equation of state at zero temperature}

At sub-MeV temperatures, the EOS is not very sensitive to variation in the
temperature. In our calculations, the $T$ = 0 EOS is obtained by
quadratic extrapolation from the results at the two lowest
temperatures $T$ = 0.158 and 0.251 MeV \cite{Lattimer09}.
Specifically, thermodynamic quantities (such as the free energy) have approximate temperature dependence,
\beq\label{T=0} F(T) = F(T=0) + a\cdot T^2, \eeq
and this equation is used to find $F(T=0)$.

%{\it T = 0 part interpolation} 
We  start with the zero temperature $T=0$ part of the EOS that constitutes the most important contribution to the adiabatic index $\Gamma$ and the speed of sound, particularly at low temperature.  The latter two quantities mostly rely on the derivative $(dP/dn)|_T$, which involves a second derivative of free energy with respect to density and this is sensitive to noise in the bicubic interpolation scheme that we use.  We start from the original Table \ref{tab:phasespace}, which is used to generate the pressure at zero T (we also include the electron pressure in the following smoothing procedure).  The thermodynamic pressure $P_{th}$ can be obtained numerically from the free
energy per baryon $F/A$, 
\beq
\label{fpressure} P_{th}\ =\ n_b^2
\left(\frac{\partial (F/A)}{\partial n_b}\right)_{T,Y_p}. 
\eeq
We smooth the pressure in a 10 points per decade table by removing points that differ significantly from the geometric mean of neighboring points and replace them with interpolated values.  Then we use monotonic cubic Hermite interpolation \cite{COtt} on the smoothed table of pressure to get a finer table with 40 points pre decade in the density axis. Finally we obtain the (free) energy at zero temperature by integrating this pressure with respect to density.

\subsection{The entropy at finite temperature}
%\subsection{Bicubic Hermite interpolation}

%{\it Monotonicity of energy and entropy in T direction} 

Next, we calculate the entropy at finite temperatures $T<12.5$ MeV.
From the free energy points at temperature $T$, density $n_B$, and proton fraction $Y_P$ as in Table~\ref{tab:phasespace}, the entropy per baryon $s_{th}$ can be obtained numerically,
\beq\label{fentropy} s_{th}\ \equiv\ S/A\ =\ -\left(\frac {\partial (F/A)}{\partial
T}\right)_{n_b, Y_p}. \eeq 
Finally, the energy per baryon $e_{th}$ is
\beq
\label{fenergy} e_{th} = F/A - T s_{th}.
\eeq
Note that for higher temperatures $T \geq$ 12.5 MeV, matter is uniform for all proton fractions and densities.  For uniform matter, all thermodynamic quantities can be obtained directly from relativistic mean field calculations, with good thermodynamic consistency.    

The bicubic interpolation scheme we use does not guarantee the monotonicity of the second derivative, {\it e.g.} 
\beq 
\frac{\partial S}{\partial T}\ =\ -\frac{\partial^2 F}{\partial^2 T} \geqslant 0. 
\eeq 
As a result, the internal energy per baryon $e_{th}$ is not guaranteed to be monotonic either, 
\beq 
\frac{\partial e_{th}}{\partial T}\ \geqslant 0. 
\eeq 
%Usually the models themselves could generate the entropy and pressure that preserve good thermodynamic consistency. Therefore one could interpolate the entropy, pressure, {\it etc} indepdently, and ensure the interpolated values preserve monotonicity. However in the microscopic mean field calculation, the entropy and/or pressure could not be calculated by any suitable analytic thermodynamic relation, but have to be obtained by numerical method as in Eqs.~(\ref{fpressure}), and (\ref{fentropy}). 
The bicubic interpolation with slope limiter can guarantee that the free energy is monotonic and that the first law of thermodynamics will be satisfied. But it is not sufficient to preserve monotonicities of the entropy and internal energy, and it can not guarantee smoothness of $(dP/dn)|_T$ as well.

The following scheme is used to generate a big table of EOS that is thermodynamically consistent, keeping monotonicities of entropy and energy, and perserving the smoothness of $(dP/dn)|_T$. We start with the entropy in the ($T, n_b$) plane as in Table \ref{tab:phasespace} with ten points per decade of density and temperature. Then we sweep through the raw table of entropy, discarding numerically noisy points that violate the following constraints:
\beq
\label{ds} 
\frac{\partial S}{\partial T} > 0,\ \frac{\partial S}{\partial n} < 0. 
\eeq 
We replace those points by interpolation from neighboring points. Therefore we have a table of entropy that satisfies conditions (\ref{ds}). (The electron and photon parts can be handled more accurately, by monotonic interpolation on individual quantities like energy, pressure and entropy.) Then we further smooth the entropy as a function of density while keeping (\ref{ds}), by further replacing noisy points with interpolated values.  This procedure ensures the smoothness of $(dP/dn)|_T$. 

Now we perform monotonic cubic Hermite interpolation on this small table of entropy with ten points per decade, to generate a larger EOS table with 40 points per decade in both the temperature and density directions, as indicated in Table \ref{tab:phasespace2}. Then we integrate this smoothed entropy table as a function of temperature to get values of the free energy (adding the energy at zero temperature to make the full free energy). Thus we obtain the free energy on this finer 40 by 40 points per decade grid in a thermodynamically consistent fashion. 

\begin{table}[h]
\centering \caption{Range of temperature $T$, density $n_B$, and proton
fraction $Y_P$ in the finely spaced interpolated EOS table.} \label{tab:phasespace2}\btab{cccc}
\hline \hline
Parameter & minimum & maximum & number of points \\
 \hline
T [MeV] & 0, 10$^{-0.8}$ & 10$^{1.9}$ & 110 \\

log$_{10}$(n$_B$) [fm$^{-3}$] &-8.0 & 0.2 & 329 \\

Y$_P$  & 0, 0.05  & 0.56 & 1(Y$_P$=0)+52 \\

\hline \etab
\end{table}

\subsection{Bicubic Interpolation of F/A} 

The final step is to carry out bicubic interpolation of the previous free energy values (as in Table \ref{tab:phasespace2}) to generate the entropy and pressure by thermodynamic derivatives Eqs.~(\ref{fpressure},\ref{fentropy}). This prescription guarantees the monotonicity of entropy and pressure in the final table, and conserves the first law of thermodynamics in adiabatic compression tests.

We first apply bicubic interpolation \cite{numericalrecipe} for the free energy.  
The first derivatives on the grid points are generated from monotonic cubic Hermite interpolation \cite{COtt}. The second derivative - the cross derivative $\partial^2 F/\partial n \partial T$, on the grid points is generated as in Ref.~\cite{numericalrecipe}. The bicubic interpolation can then fit free energies with cubic functions in temperature and density coordinates, and provides the first and second (cross) derivatives. Then the entropy and pressure are otained from Eqs.~(\ref{fpressure},\ref{fentropy}). Finally, the boundary points at the highest temperature or density are discarded to avoid boundary artifacts in the interpolation.

\ \

\section{\label{result}Results for the equation of state}

In this section we discuss various thermodynamic quantities and the composition in the EOS. First, we discuss the free energy per baryon from Virial
gas, nonuniform Hartree mean field, and uniform matter calculations.  Next we map
out the phase boundaries of our EOS. Then we show
various thermodynamic quantities such as the pressure, entropy per baryon, chemical potentials of neutrons, protons, and electron neutrinos, and the
adiabatic index. We also show information on the composition, 
{\it i.e.}, the average mass number and
proton number of heavy nuclei in non-uniform matter, and the mass
fractions of different species. The zero temperature EOS in
chemical equilibrium is presented next, and this is used to
calculate neutron star structure.  Finally, we make some comparisons
between our EOS and those of Lattimer-Swesty \cite{LS} and H. Shen
\etal \cite{Shen98}.

\subsection{Free energy and phase boundaries}

\begin{widetext}

\begin{figure}[htbp]
 \centering
 \includegraphics[height=8.5cm,angle=-90]{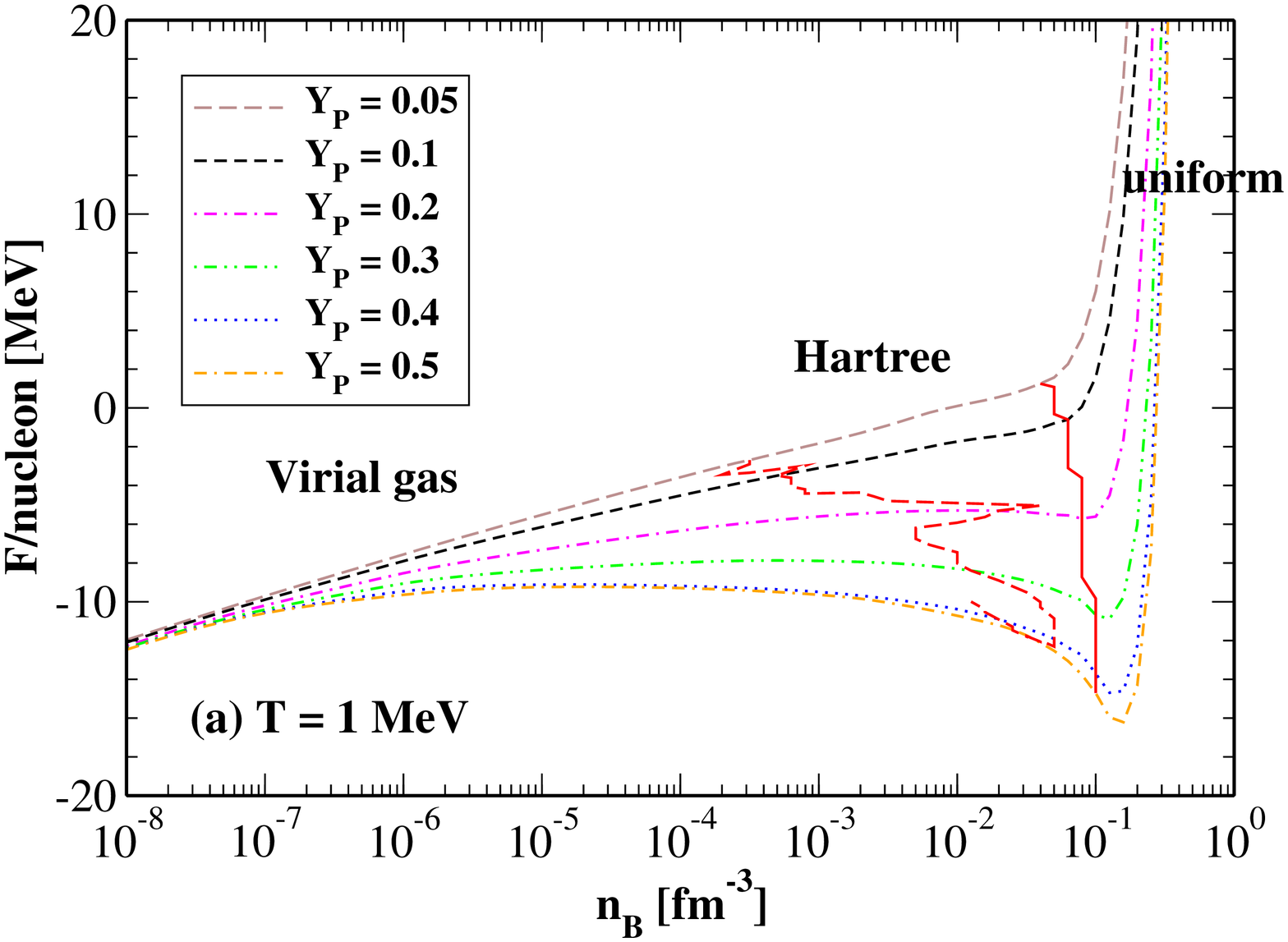}
 \includegraphics[height=8.5cm,angle=-90]{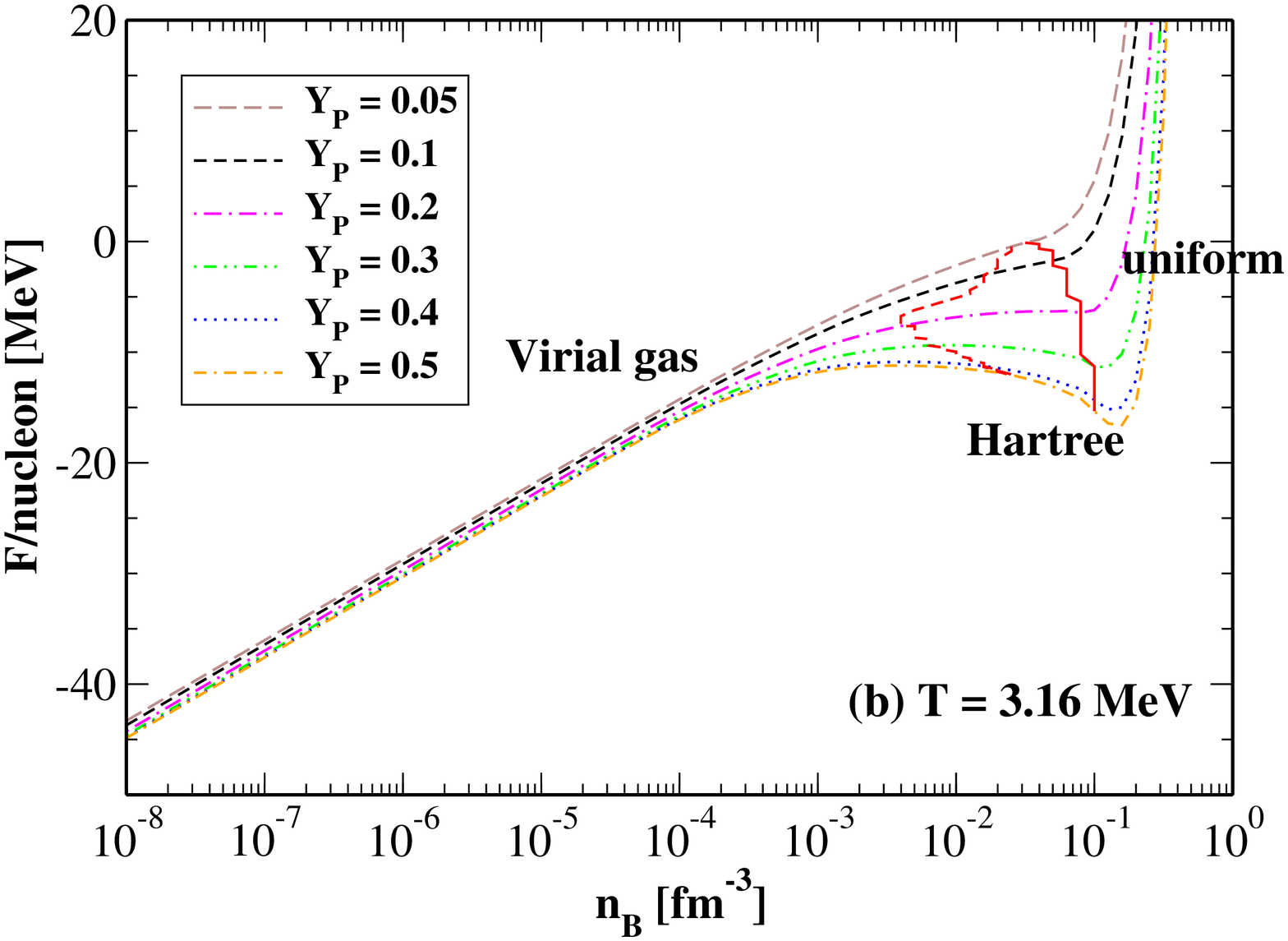}
 \includegraphics[height=8.5cm,angle=-90]{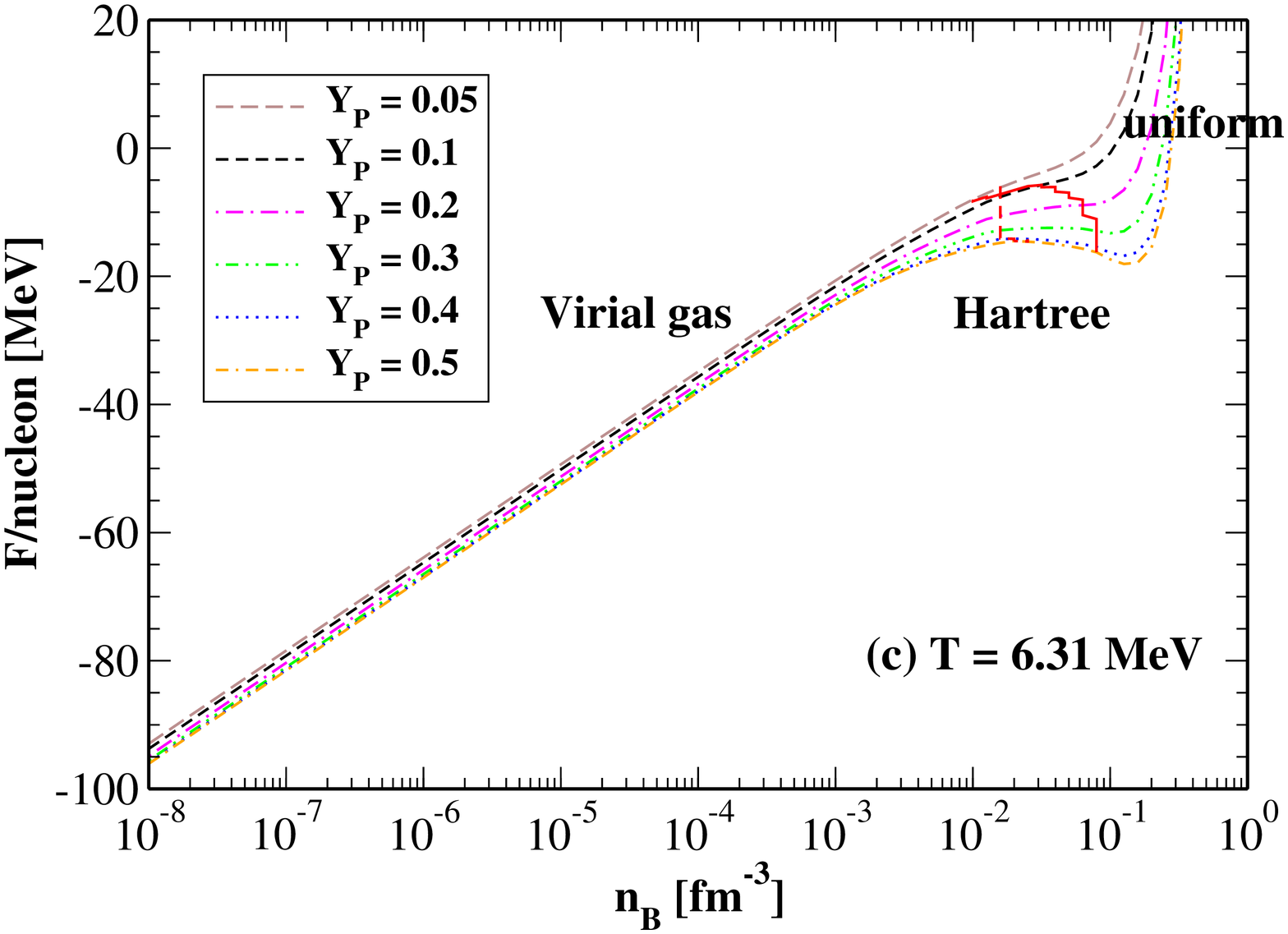}
 \includegraphics[height=8.5cm,angle=-90]{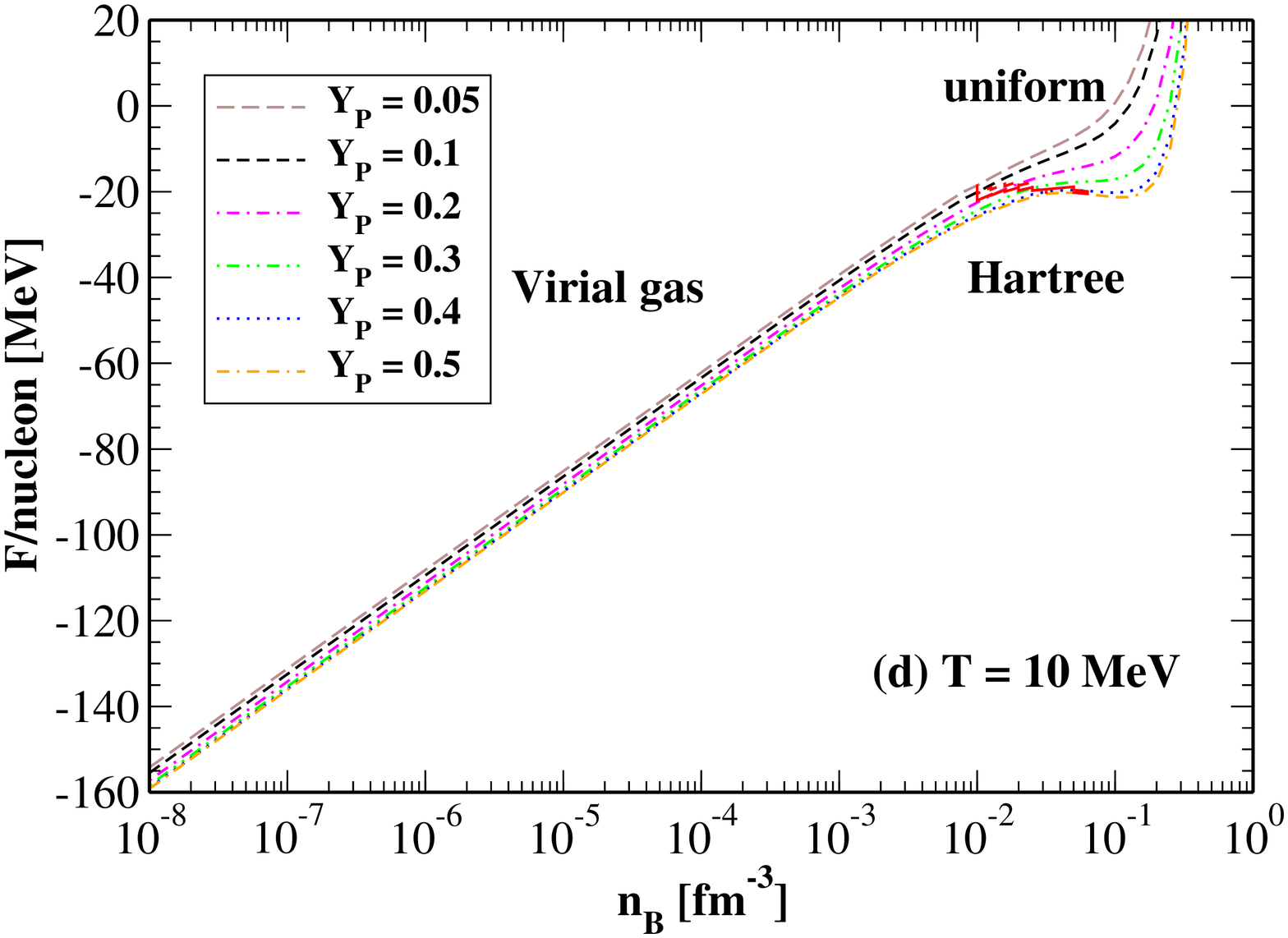}

\caption{(Color online) Free energy per nucleon of nuclear matter at temperatures of T = 1 (a), 3.16 (b), 6.31 (c), and 10 (d) MeV. The proton fraction ranges from Yp = 0.05 to 0.5. }\label{fig:fe}
\end{figure}

\end{widetext}

In Fig.~\ref{fig:fe}, the free energy per baryon $F/A$ for matter at
$T$ = 1, 3.16, 6.31 and 10 MeV with different proton fractions is
shown as a function of density $n_B$.  The free energy $F/A$ is obtained from
Virial gas, nonuniform Hartree mean field, and
uniform matter calculations, respectively.  In most cases
the transition (as the density grows) is found at the density where
Hartree calculations or uniform matter calculations give the lowest free energy.  For
matter at very low temperature (not greater than $\sim$ 1 MeV) and
low proton fraction (not greater than $\sim$ 0.1), some matching points are obtained at densities where Virial gas and Hartree calculations
give the closest free energies (the difference is less than hundreds of
KeV, and note that that mass table \cite{FRDM} used in our Virial EOS \cite{SHT10b} has itself 600 KeV rms deviations in nuclear binding energies compared to data).   Detailed information has been reported in our previous paper \cite{SHT10b}.  We reproduce the figure here for completeness.

\begin{widetext}

\begin{figure}[htbp]
 \centering
 \includegraphics[height=13cm,angle=-90]{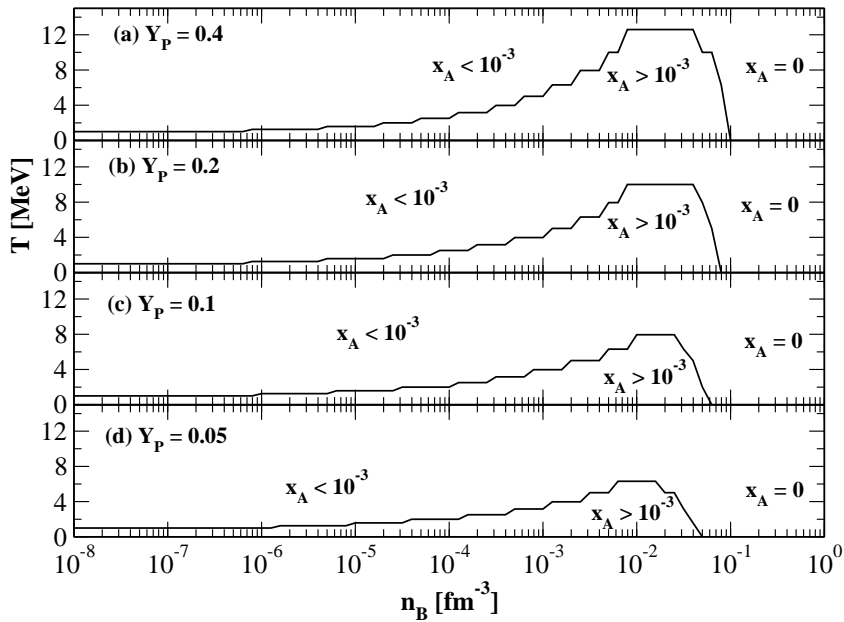}

\caption{Phase diagram of nuclear matter at different proton
fractions, (a) 0.4, (b) 0.2, (c) 0.1 and (d) 0.05. $X_A$ is mass fraction of heavy
nuclei with mass number $A$ $>$ 4.}\label{fig:phaseboundary}
\end{figure}

\end{widetext}

In Fig.~\ref{fig:phaseboundary}, we show the phase boundaries of
nuclear matter at different proton fractions $Y_P$ = 0.05, 0.1,
0.2 and 0.4. The mass fraction of heavy nuclei with mass number $A>4$ is $X_A$.  The boundary at low densities indicates when $X_A$ is greater or
less than 10$^{-3}$. The boundary at high densities indicates the
transition between non-uniform matter and uniform matter.  At very
low densities, the matter is dominated by free nucleons and alpha
particles.  As the density rises, heavy nuclei persist to higher temperatures.  Finally uniform matter takes over at sufficiently high density.  Fig.~\ref{fig:phaseboundary} also shows that as proton fraction rises, the temperature regime with
appreciable heavy nuclei grows and the transition density to
uniform matter increases.  The density for the nonuniform
matter to uniform matter transition has a weak temperature dependence.
Fig.~\ref{fig:phaseboundary2} shows the phase diagram of nuclear
matter at different temperatures $T$ = 1, 3.16, 6.31, and 10 MeV.
Similarly as in Fig.~\ref{fig:phaseboundary}, the left boundary at low density
indicates where $X_A$ is greater or less than 10$^{-3}$. The right
boundary at high density indicates the transition between
non-uniform matter and uniform matter.  As the temperature grows, the
$Y_P$ dependence of the phase boundaries becomes larger, and the
density regime with appreciable heavy nuclei rapidly shrinks. Our Figs.~\ref{fig:phaseboundary} and \ref{fig:phaseboundary2} are very simillar to the corresponding Figs. 2 and 3 of ref. \cite{Shen98} for the H Shen \etal~ EOS.

\subsection{Pressure}

In Fig.~\ref{fig:pressure}, the pressure of nuclear matter with
different proton fractions is shown as a function of density, for
four different temperatures (1, 3.16, 6.31, and 10 MeV). The
pressure is obtained via the bicubic Hermite interpolation, in
Eq.~(\ref{fpressure}).  For matter at low density in the Virial gas
phase, the pressure scales nicely with density. When the density
increases, formation of additional nuclei decreases the pressure. The Coulomb
correction to the lattice energy reduces the pressure even further.  Note in part of
Virial gas and Hartree Wigner-Seitz cell regions, the pressure is negative.  Figure ~\ref{fig:pressure} only shows hadronic contribution to the pressure.  In addition one needs to take into account the contribution from electrons and photons, which makes the total pressure (not shown here) positive.  We also note that with inclusion of electrons and photons in the final table, the condition for thermodynamic stability against density fluctuations is always preserved to the numerical precision of the table, \beq \frac{dP}{dn}\mid_T \geq 0. \eeq  Finally after the Hartree to uniform matter transition, the pressure rises rapidly with density because of the large incompressibility of nuclear mater.

\subsection{Entropy}
In the upper left panel of Fig.~\ref{fig:s}, the entropy per baryon is
shown as a function of density for nuclear matter with $T$ = 1 MeV
at different proton fractions.  The entropy is calculated from
Eq.~(\ref{fentropy}).  Near a density of a few times 10$^{-8}$ fm$^{-3}$, the
entropy drops rapidly with density, which indicates formation of
nuclei, first mostly alpha particles and then heavier
nuclei at higher density.    Later we will illustrate this by plotting the mass fractions of different species.  For nuclear matter with a large proton fraction,
larger mass fractions of heavy nuclei are present so that the entropy drops more rapidly with density.  When nuclei compose more than 90\% of the mass fraction, the entropy remains almost constant as the density increases. The other panels in
Fig.~\ref{fig:s} show the entropy per baryon in nuclear matter at
higher temperatures 3.16, 6.31, and 10 MeV. They have similar
characteristics as that for T = 1 MeV, except that the range of
densities where the entropy drops rapidly, due to formation of nuclei,
becomes smaller because it is harder to form heavy nuclei at higher
temperatures.

\begin{widetext}

\begin{figure}[htbp]
 \centering

 \includegraphics[height=13cm,angle=-90]{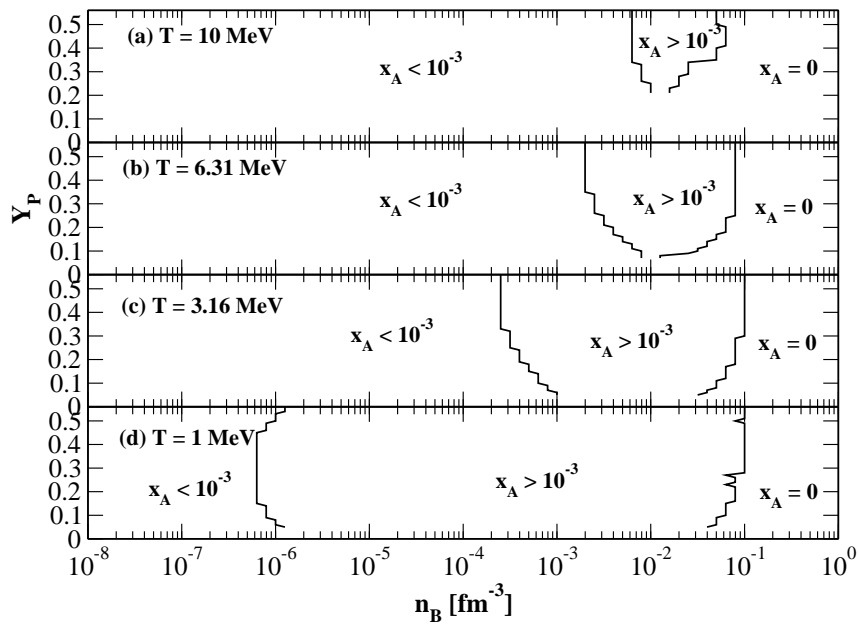}

\caption{Phase diagram of nuclear matter at different
temperatures, (a) 10, (b) 6.31, (c) 3.16 and (d) 1 MeV. $X_A$ is the same as
Fig.~\ref{fig:phaseboundary}.}\label{fig:phaseboundary2}
\end{figure}

%\end{widetext}
%\begin{widetext}

\begin{figure}[htbp]
 \centering
 \includegraphics[height=8.5cm,angle=-90]{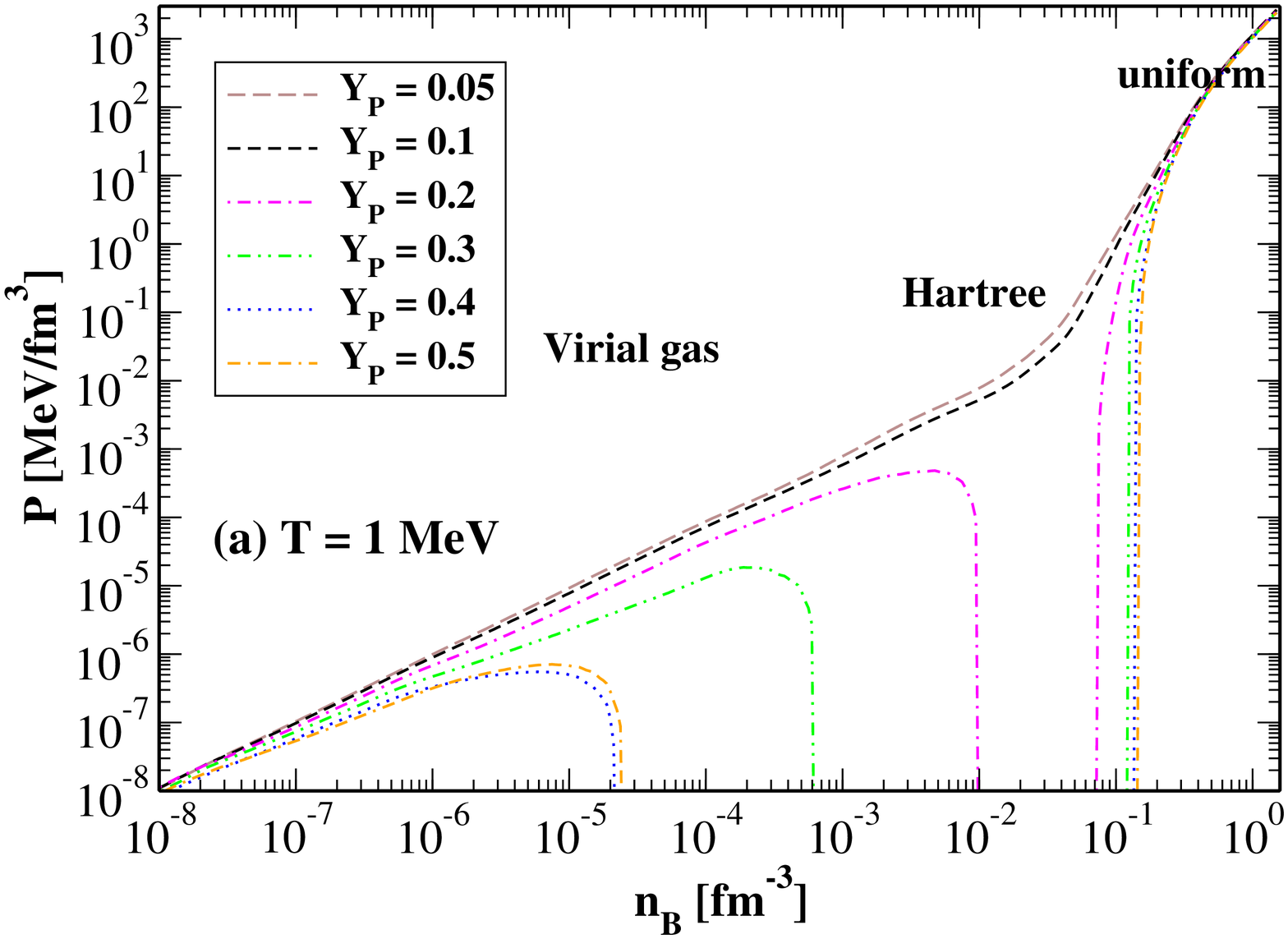}
 \includegraphics[height=8.5cm,angle=-90]{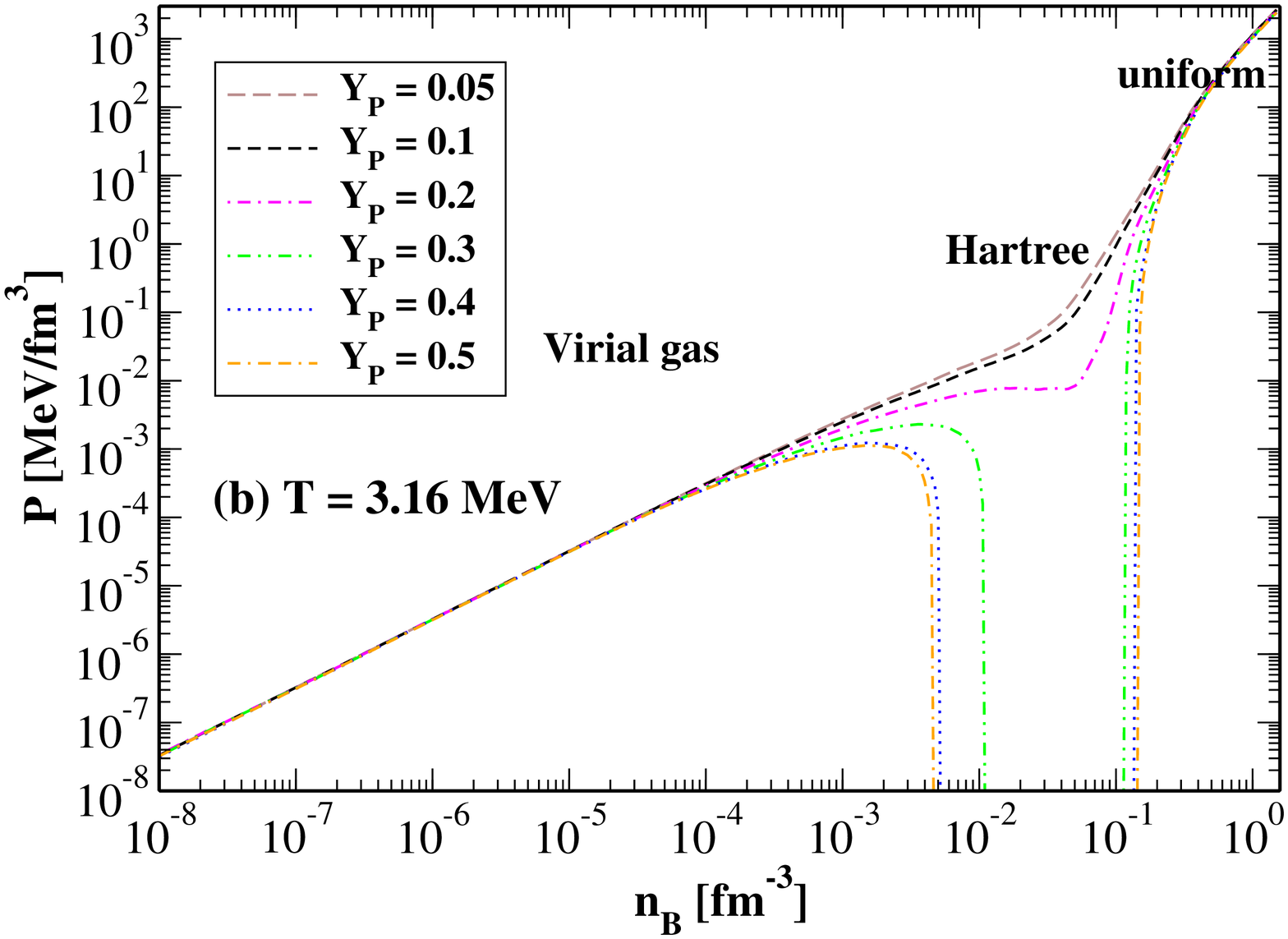}
 \includegraphics[height=8.5cm,angle=-90]{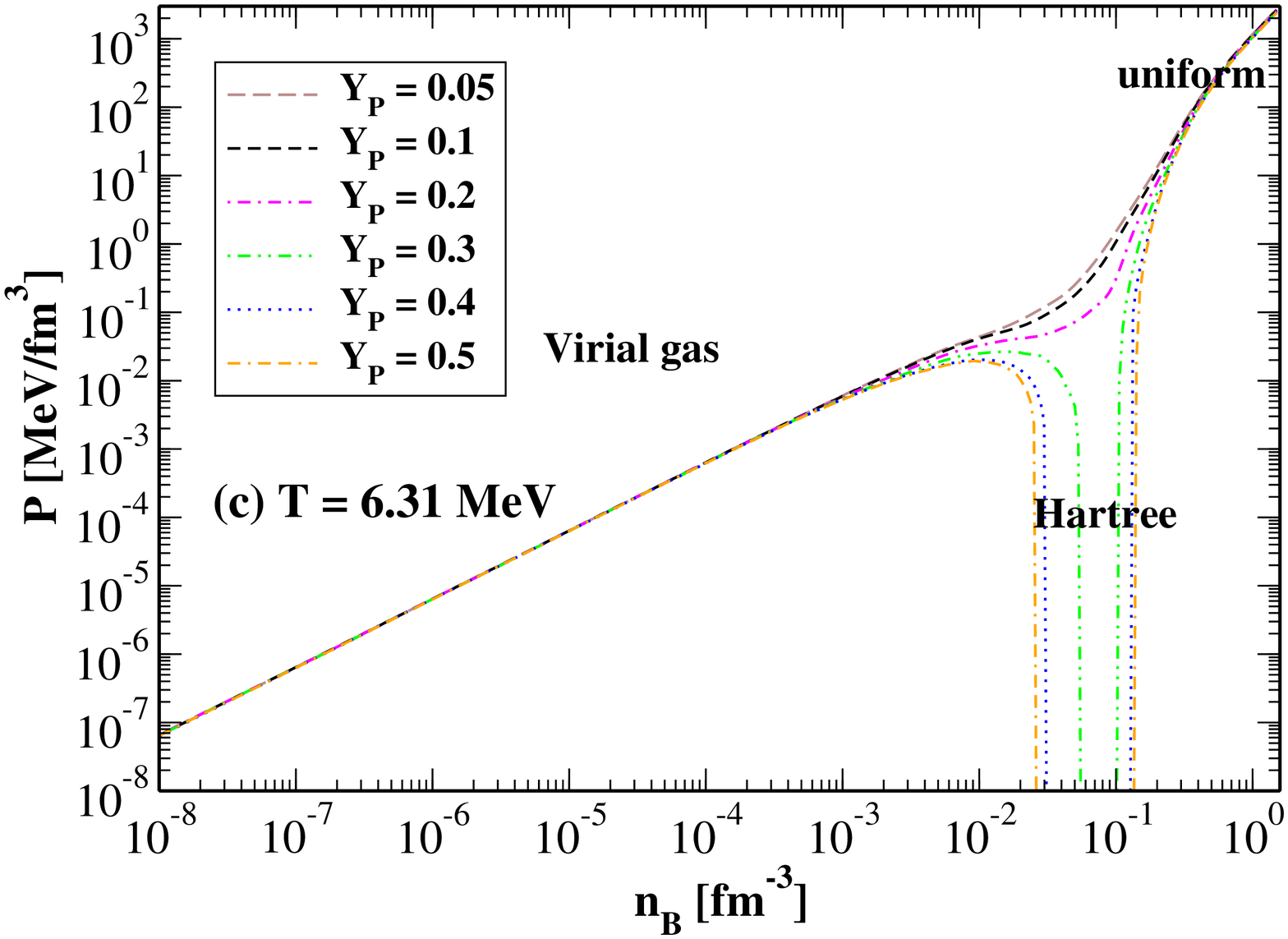}
 \includegraphics[height=8.5cm,angle=-90]{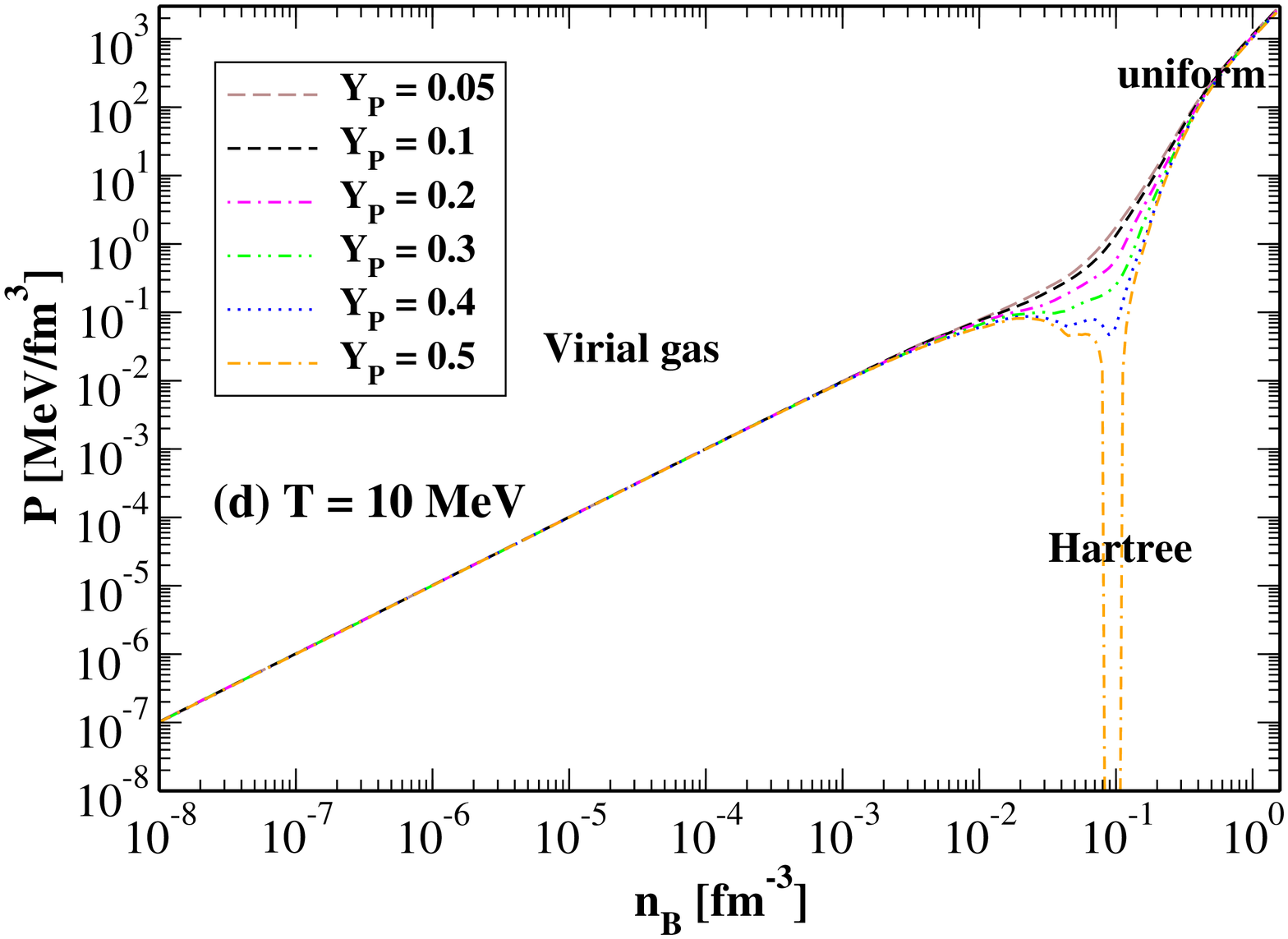}

\caption{(Color online) Pressure of nuclear matter at temperatures of T = 1 (a), 3.16 (b), 6.31 (c), and 10 (d) MeV. The proton fraction ranges from Yp = 0.05 to 0.5.}\label{fig:pressure}
\end{figure}

%\end{widetext}

%\begin{widetext}

\begin{figure}[htbp]
 \centering
 \includegraphics[height=8.5cm,angle=-90]{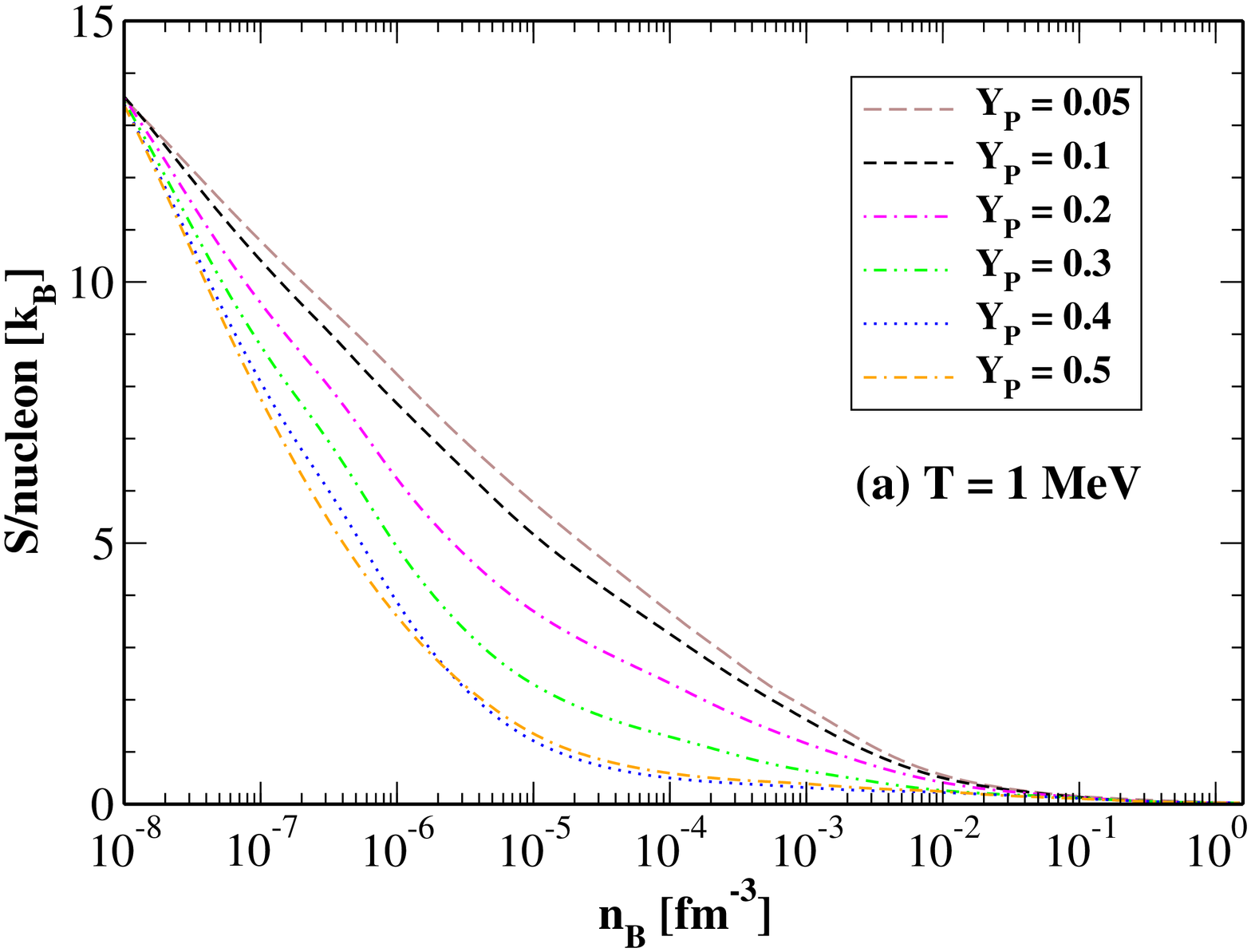}
 \includegraphics[height=8.5cm,angle=-90]{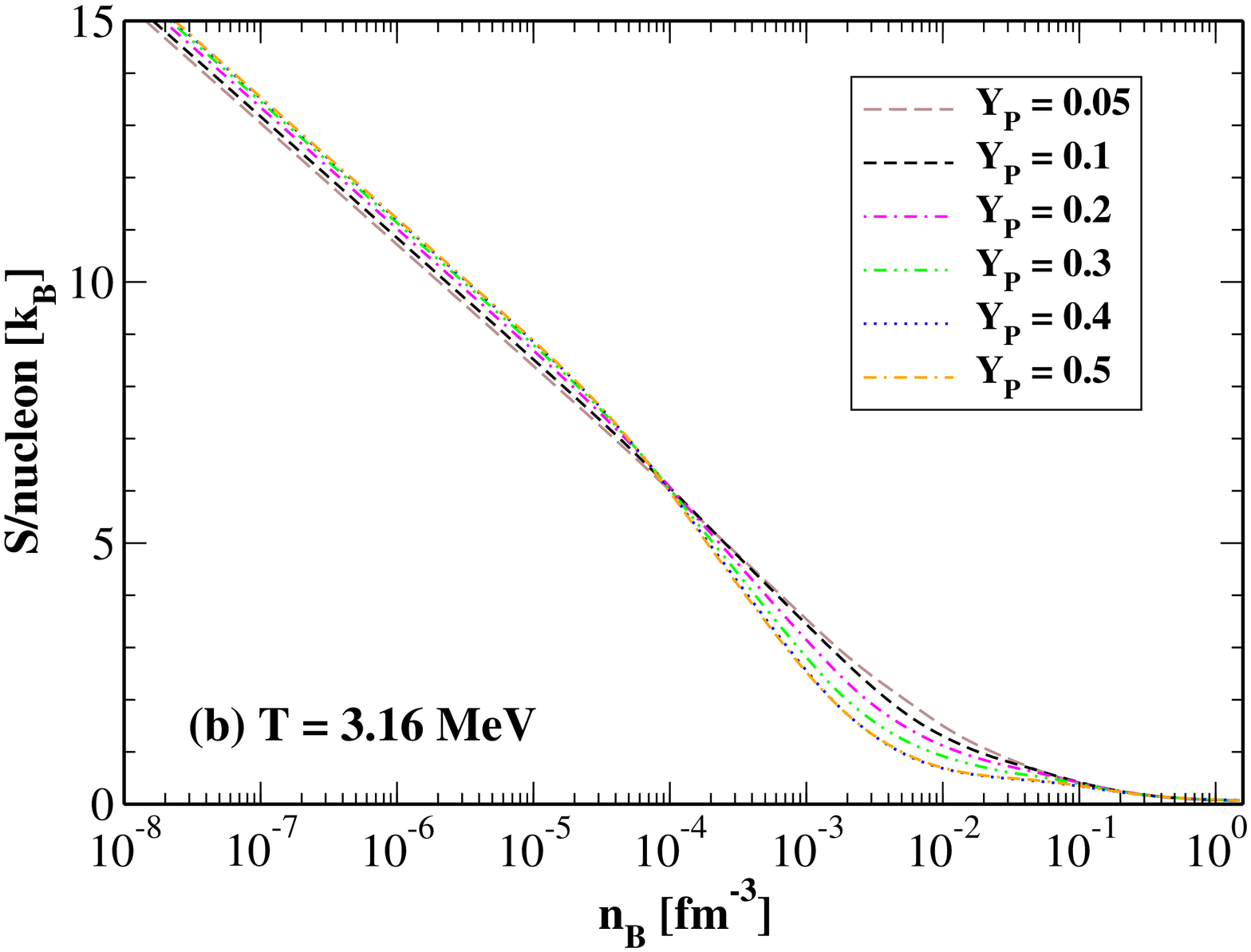}
 \includegraphics[height=8.5cm,angle=-90]{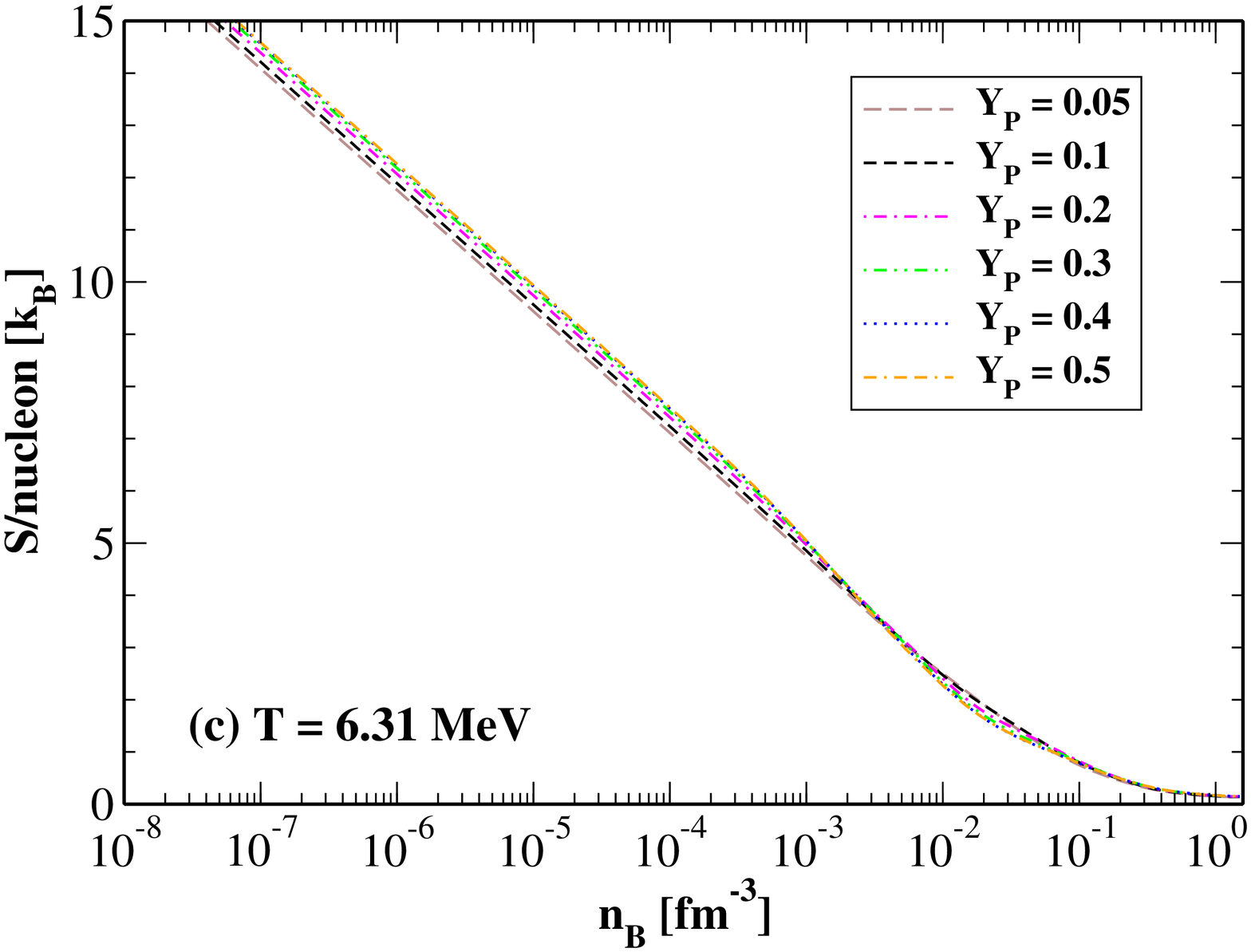}
 \includegraphics[height=8.5cm,angle=-90]{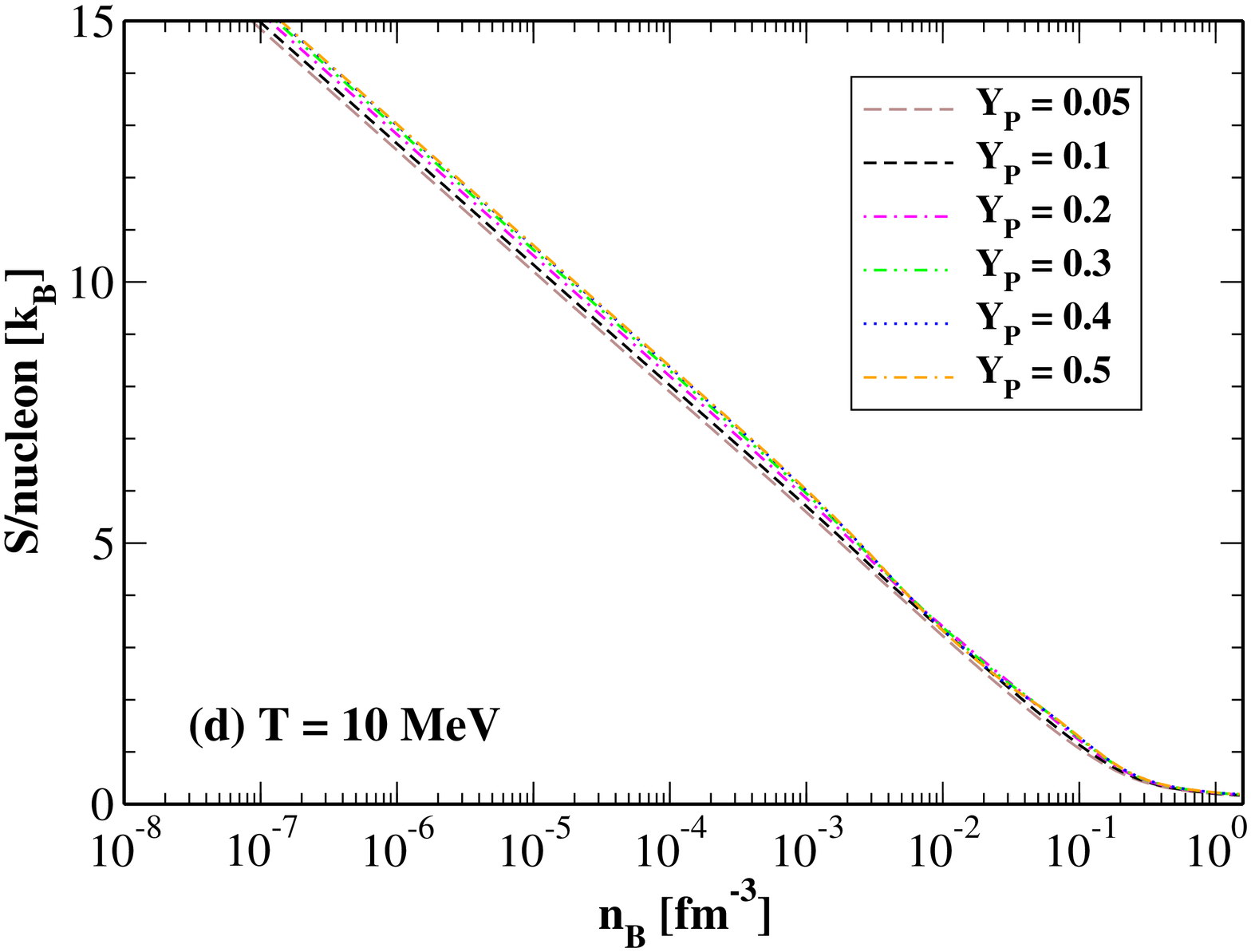}

\caption{(Color online) Entropy per nucleon of nuclear matter at temperatures of T = 1 (a), 3.16 (b), 6.31 (c), and 10 (d) MeV. The proton fraction ranges from Yp = 0.05 to 0.5. }\label{fig:s}
\end{figure}

%\end{widetext}

%\begin{widetext}

\begin{figure}[htbp]
 \centering
 \includegraphics[height=8.5cm,angle=-90]{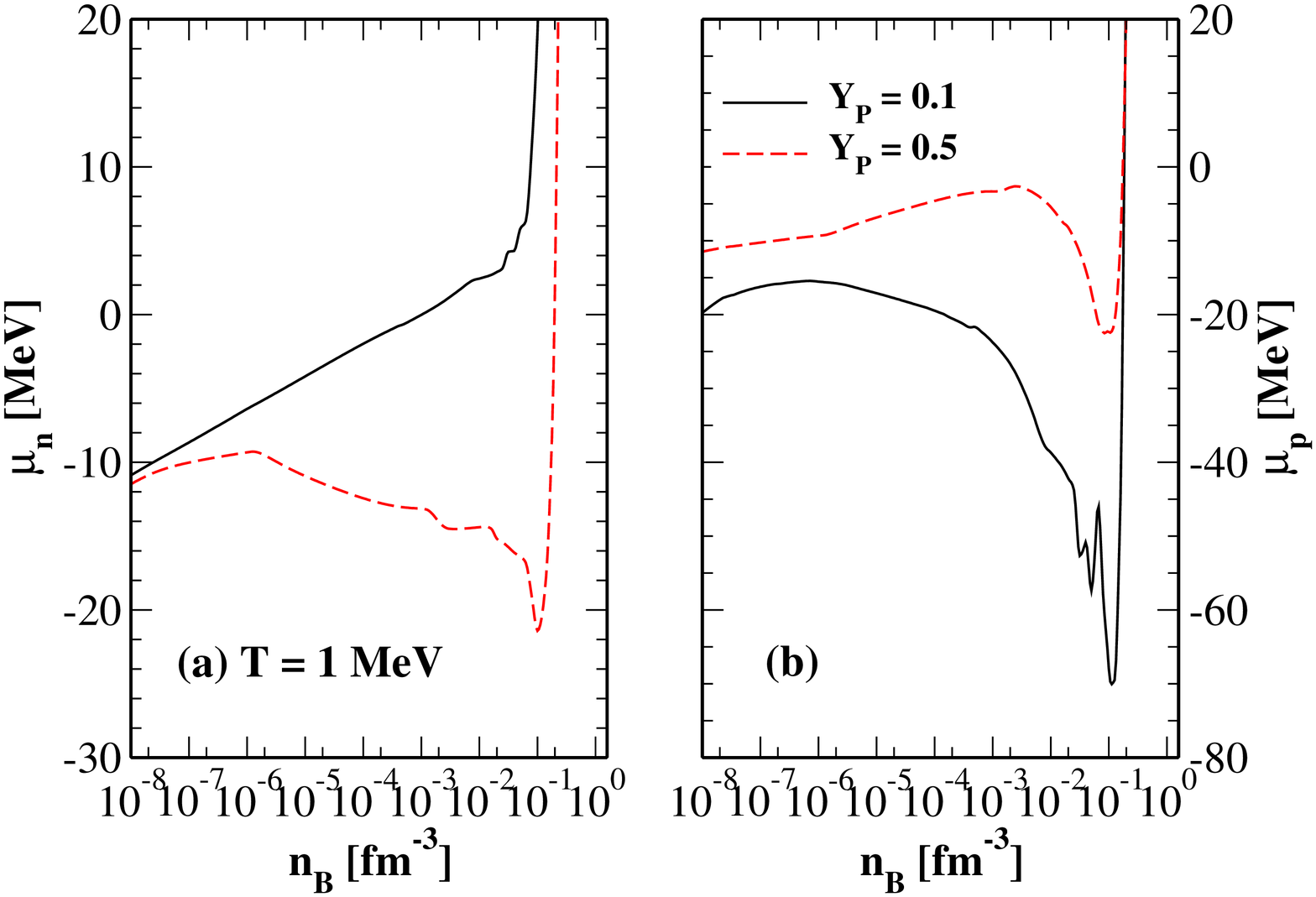}
 \includegraphics[height=8.5cm,angle=-90]{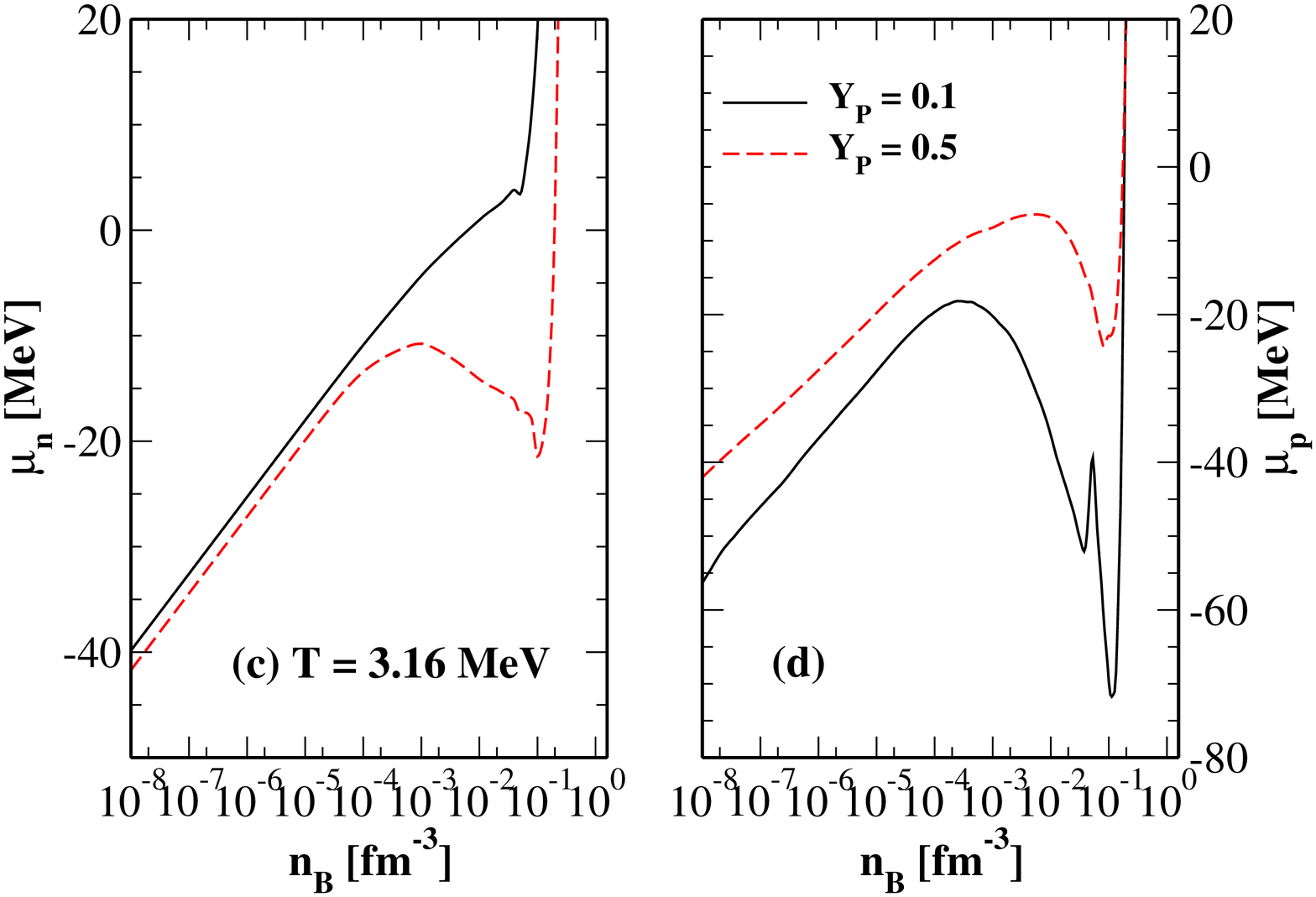}
 \includegraphics[height=8.5cm,angle=-90]{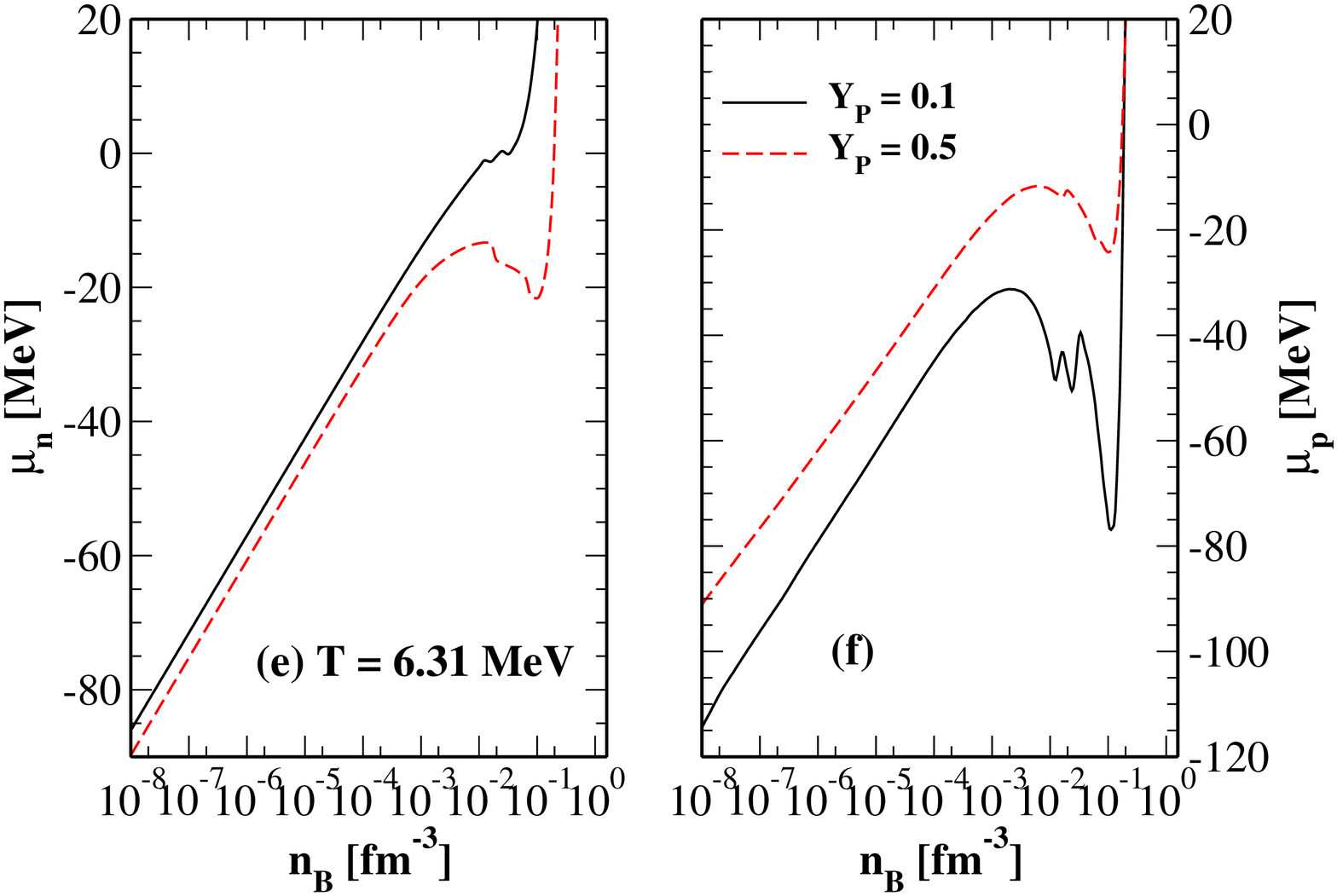}
 \includegraphics[height=8.5cm,angle=-90]{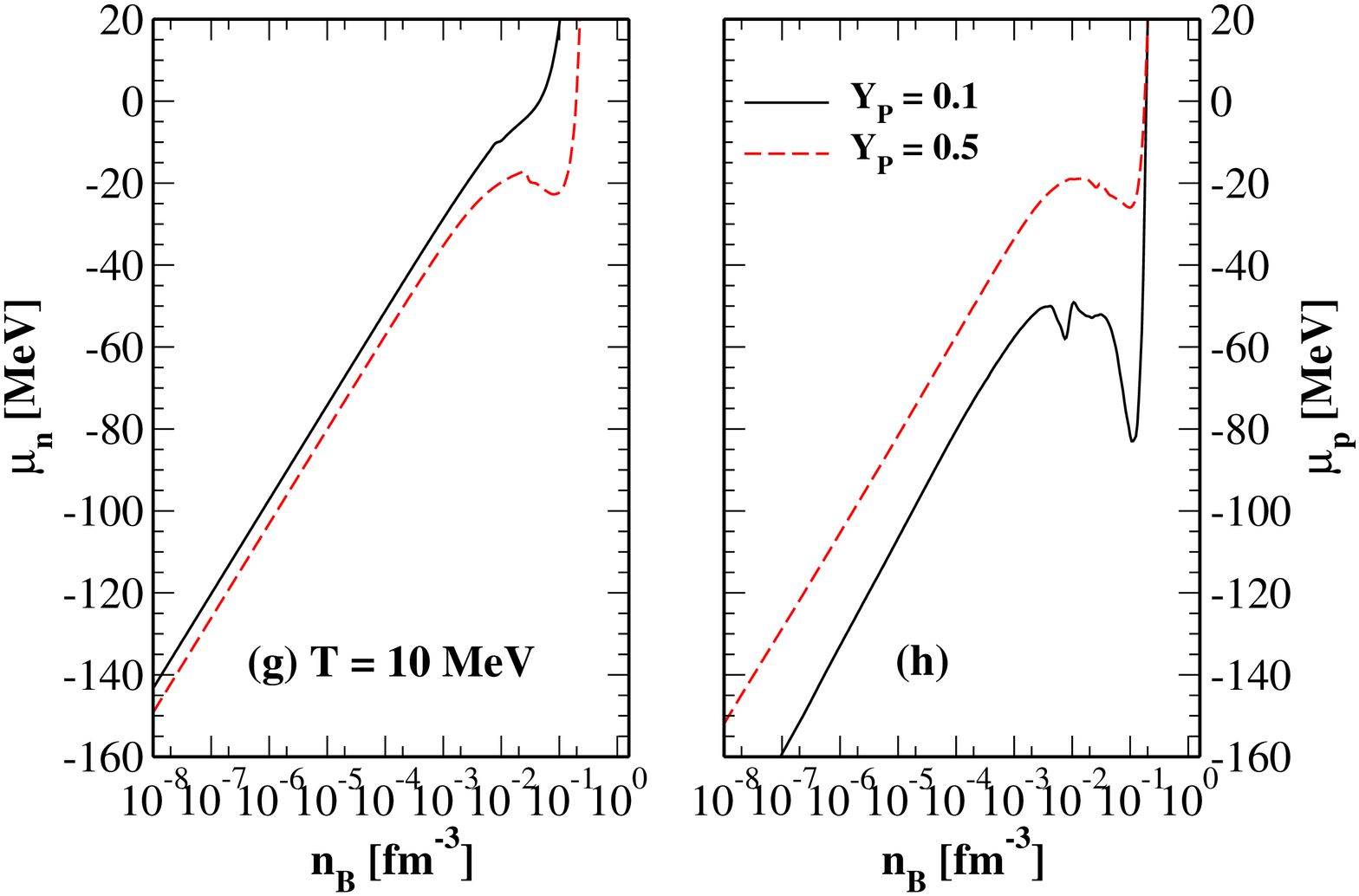}

\caption{(Color online) Chemical potentials of neutron and proton at different
temperature and proton fractions.}\label{fig:chemical}
\end{figure}

\end{widetext}

\subsection{Chemical potentials of $n$, $p$, and $\nu_e$.}

In Fig.~\ref{fig:chemical}, the neutron (left subpanels, a, c, e, g) and proton (right subpanels, b, d, f, h) chemical potentials are shown for nuclear matter with $Y_p$ = 0.1 (black curves) and 0.5 (red
curves) at different temperatures. For nuclear matter at low density
and high temperature, the chemical potentials scale with
density. The formation of nuclei breaks the scaling behavior, {\it i.e.}, the bound nucleons
have much smaller chemical potentials.

In Fig.~\ref{fig:neutrino}, the equilibrium electron neutrino chemical
potential, defined as $\mu_{\nu_e}\equiv\mu_e+\mu_p-\mu_n$, is
shown versus density for nuclear matter at various temperatures
and proton fractions.  Let us focus on the top left panel where
nuclear matter has $T$ = 1 MeV.  For symmetric nuclear matter,
$\mu_{\nu_e}$ grows rapidly with density, since $\mu_p$ almost
cancels $\mu_n$ and $\mu_{\nu_e} \sim \mu_e \sim n_B^{\frac 1 3}$.
For proton rich nuclear matter ($Y_P>0.5$), $\mu_{\nu_e}$ will grow even faster than for symmetric nuclear matter. For neutron rich nuclear matter, $\mu_{\nu_e}$ has a
maximum value at some density.  This is due to the competition
between the electron chemical potential and the difference of the proton
and neutron chemical potentials. The latter difference between proton and neutron is negative when the proton fraction is less than 0.5.  The density where $\mu_{\nu_e}$ has a maximum decreases with decreasing proton fraction.  It is $\sim$ 0.2 fm$^{-3}$
when $Y_P$ = 0.3, and $\sim$ 0.05 fm$^{-3}$ when $Y_P$ = 0.05. In
the limit of pure neutron matter there is no maximum for
$\mu_{\nu_e}$, which is just negative $\mu_n$ by construction.

\subsection{Adiabatic index $\Gamma_s$.}

The left panel of Fig.~\ref{fig:adiabatic_gamma} shows the temperature of the adiabat with entropy $S$ = 1, for nuclear matter with proton fraction $Y_P$ = 0.1, 0.2, 0.3, and 0.4 including baryon,
electron, positron, and photon contributions to $S$.  At low density, matter is
dominated by free nucleons in the gas phase. There are lots of
kinetic degrees of freedom and one needs only a small temperature for the
matter to have $S$ = 1.  When the density rises, the formation of heavy
nuclei suppresses the entropy and it takes higher temperatures for
the matter to reach $S$ = 1.  The entropy, in this case, comes from
a fraction of free nucleons in a gas and the partial occupation of
levels in the heavy nuclei.  When uniform nuclear matter forms at high density, it takes even a higher temperature to reach the same entropy $S$ = 1 because the uniform matter is highly
degenerate and the entropy comes only from excitations above the Fermi
surfaces.

The adiabatic index $\Gamma_s$, 
\beq\label{adia} \Gamma_s\ =\ \left(
\frac{\partial \mathrm{ln}P}{\partial \mathrm{ln}n}\right)_s, 
\eeq
describes the stiffness of the EOS at constant entropy.  In the
right panel of Fig.~\ref{fig:adiabatic_gamma}, the adiabatic index is shown versus
density for nuclear matter with constant entropy $S$ = 1 and  $Y_P$ = 0.1, 0.2, 0.3, and 0.4.  At
subnuclear density, $\Gamma_s$ remains almost constant with small
variation. It then rises rapidly
at the transition from nonuniform to uniform matter.  This
characterizes the stiffening of the EOS due to the large nuclear
incompressibility at high density.

\subsection{Average $\bar{A}$ and $\bar{Z}$ of heavy nuclei.}
% mass number

Figure~\ref{fig:massnumber} shows the average mass number $\bar{A}$ of
heavy nuclei (with $A>4$) versus baryon density at different temperatures and
proton fractions.  Note that alpha particles are
not counted as heavy nuclei.  Let us first look at the upper left
panel, where $T$ = 1 MeV.  Nuclear shell effects give rise to several approximate plateaus in
$\bar{A}$ vs density for each $Y_P$, for example $\bar{A}$ = 12, $\sim$ 50, $\sim$ 80 and $\sim$ 100.   Usually $\bar{A}$ is larger in matter with smaller $Y_P$. There
are oscillations in $\bar{A}$ in the Hartree mean field regime.  This
is due to both nuclear shell effects and small errors in the free energy minimization due to our using a finite step in the Wigner Seitz cell size \cite{SHT10a}.  The average mass $\bar{A}$ can be as large as 3,000 at high density before the final transition to uniform matter.  At higher temperatures, as shown in the other panels of Fig. \ref{fig:massnumber}, $\bar{A}$ grows more rapidly to several thousand in a narrower range of density.  As a result, the
width of the plateaus in $\bar{A}$ with density become much shorter and finally vanish.

The following Fig.~\ref{fig:protonnumber} shows the average proton
number $\bar{Z}$ of heavy nuclei versus baryon density at different
temperatures and proton fractions.  The average proton number has very similar
characteristics as for the average mass number, except that the differences in $\bar{Z}$ between different $Y_P$s are smaller.

% equilibrium electron neutrino chemical potentials
\begin{widetext}

\begin{figure}[htbp]
 \centering
 \includegraphics[height=8.5cm,angle=-90]{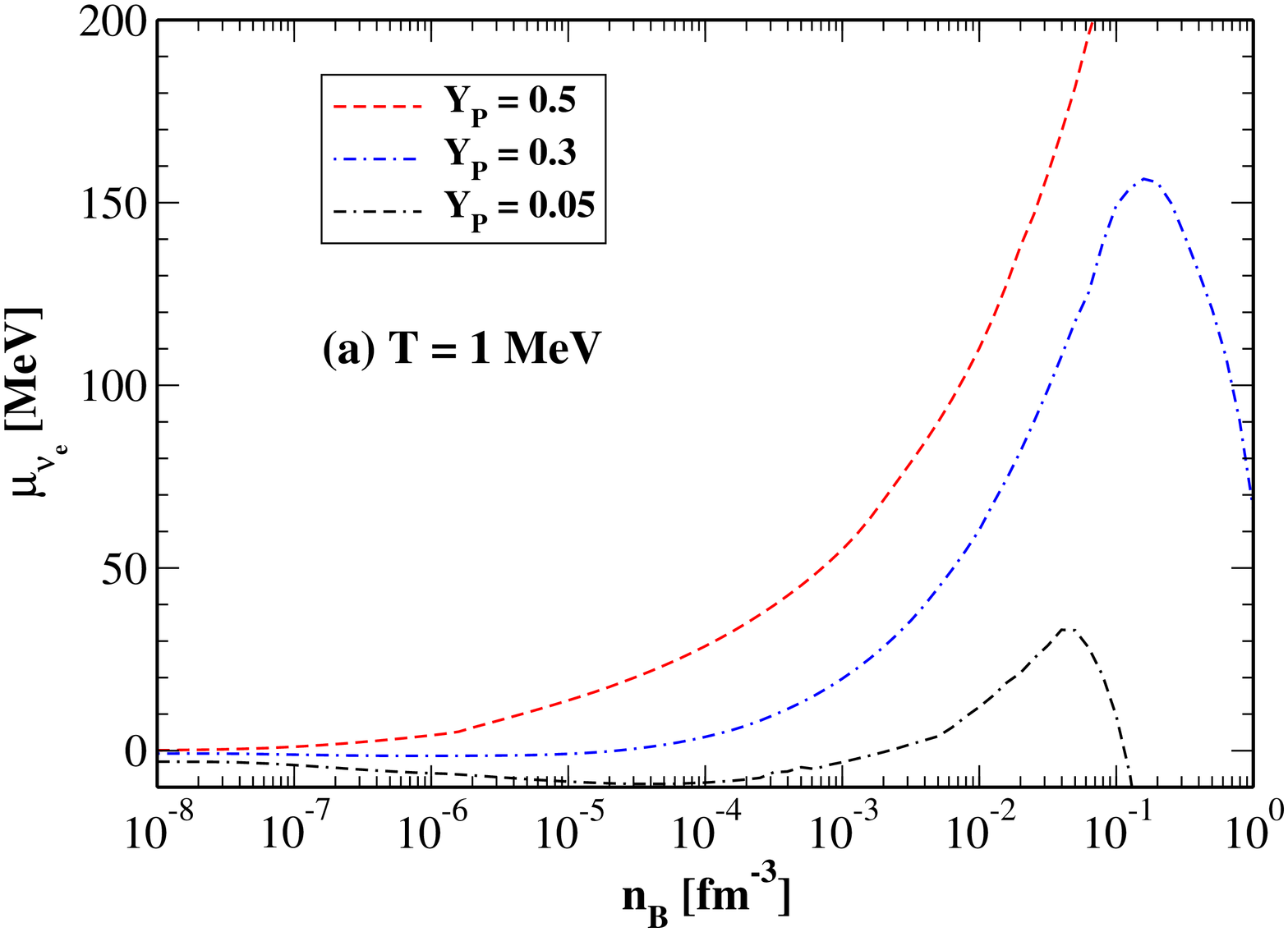}
 \includegraphics[height=8.5cm,angle=-90]{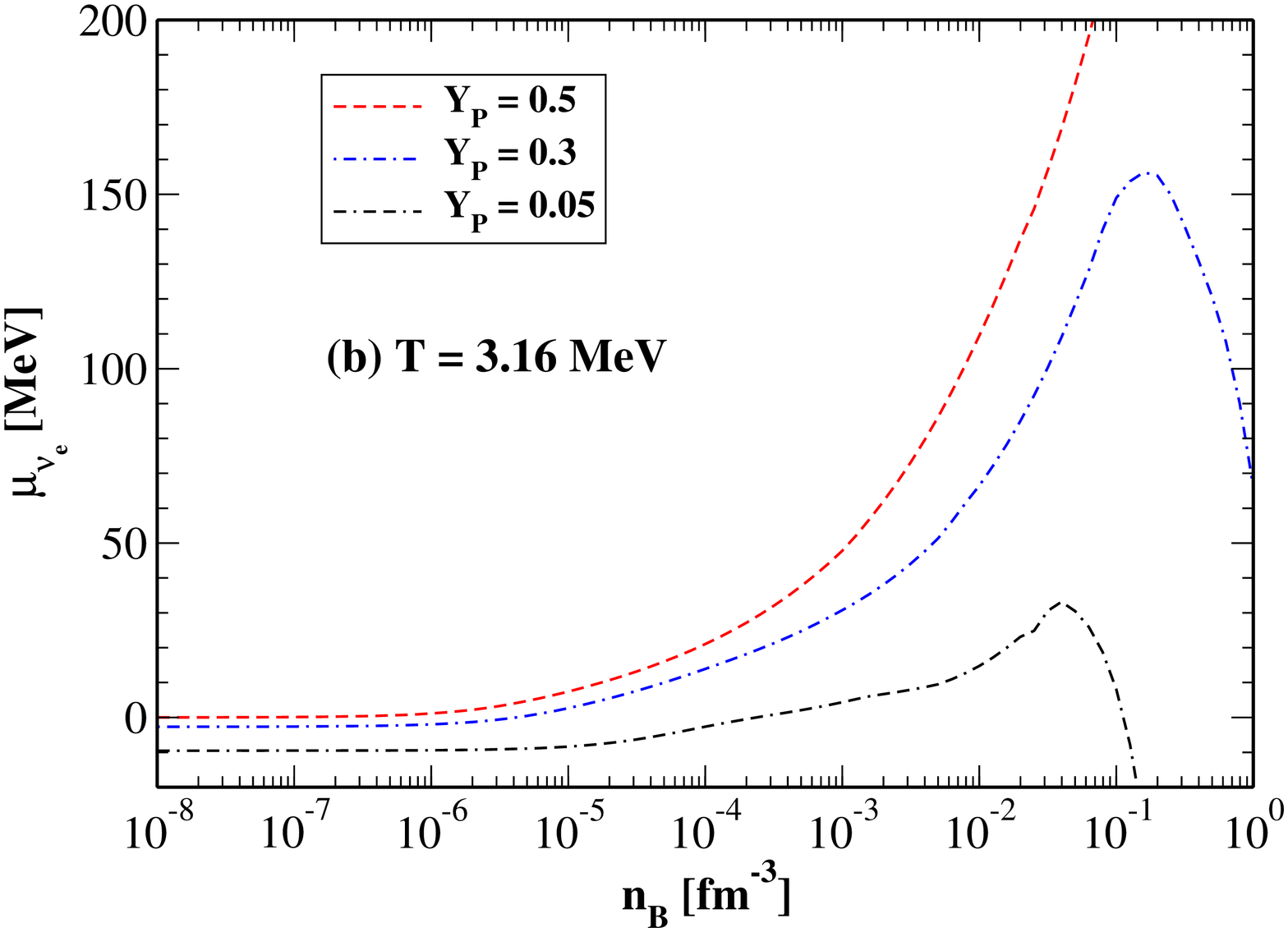}
 \includegraphics[height=8.5cm,angle=-90]{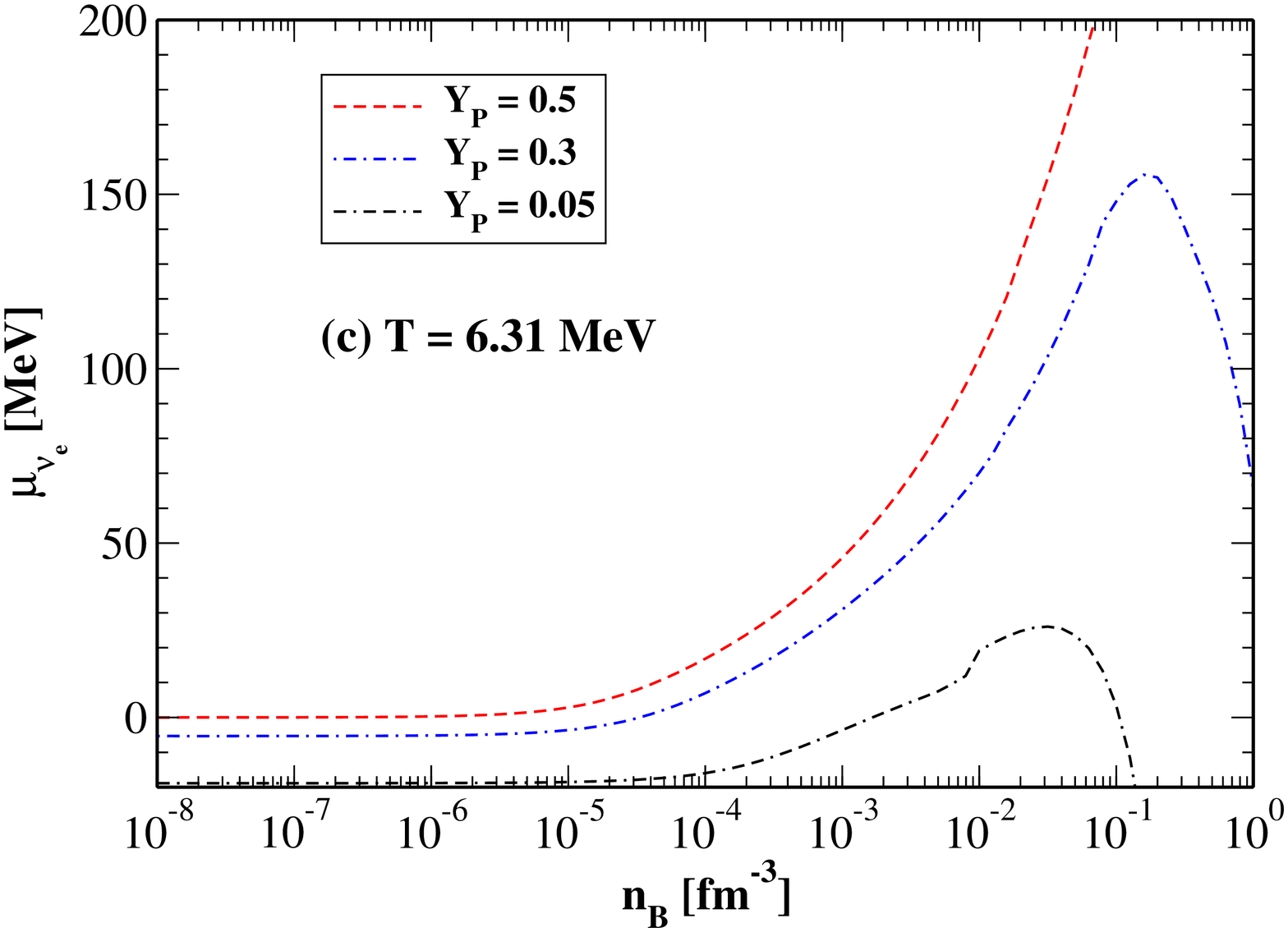}
 \includegraphics[height=8.5cm,angle=-90]{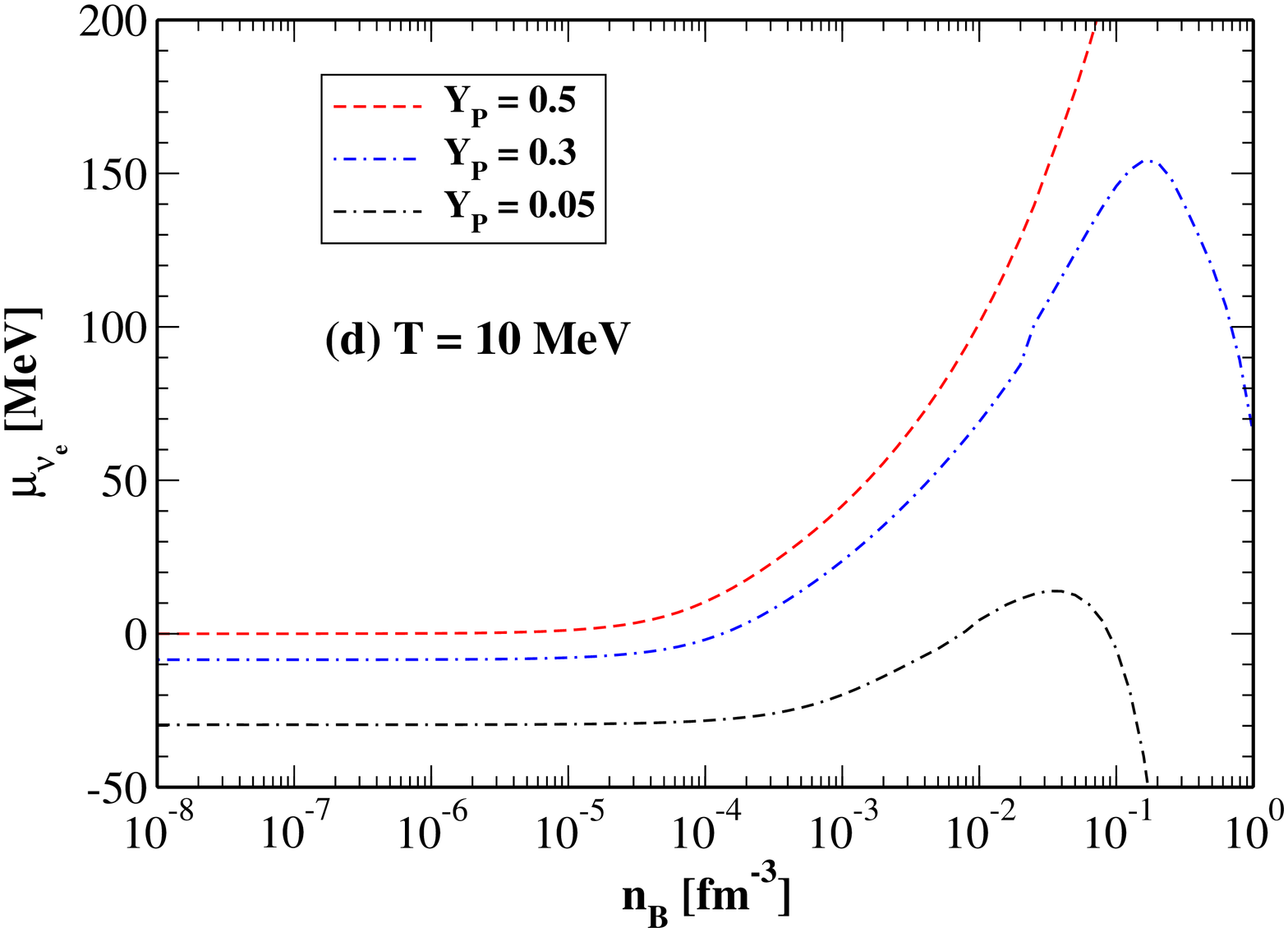}

\caption{(Color online) Equilibrium electron neutrino chemical potential $\mu_{\nu_e}$ for
nuclear matter at various temperatures and proton fractions.  It is defined as
$\mu_{\nu_e}\equiv\mu_e+\mu_p-\mu_n$.}\label{fig:neutrino}
\end{figure}

%\end{widetext}

%\begin{widetext}

\begin{figure}[htbp]
 \centering
 \includegraphics[height=8.5cm,angle=-90]{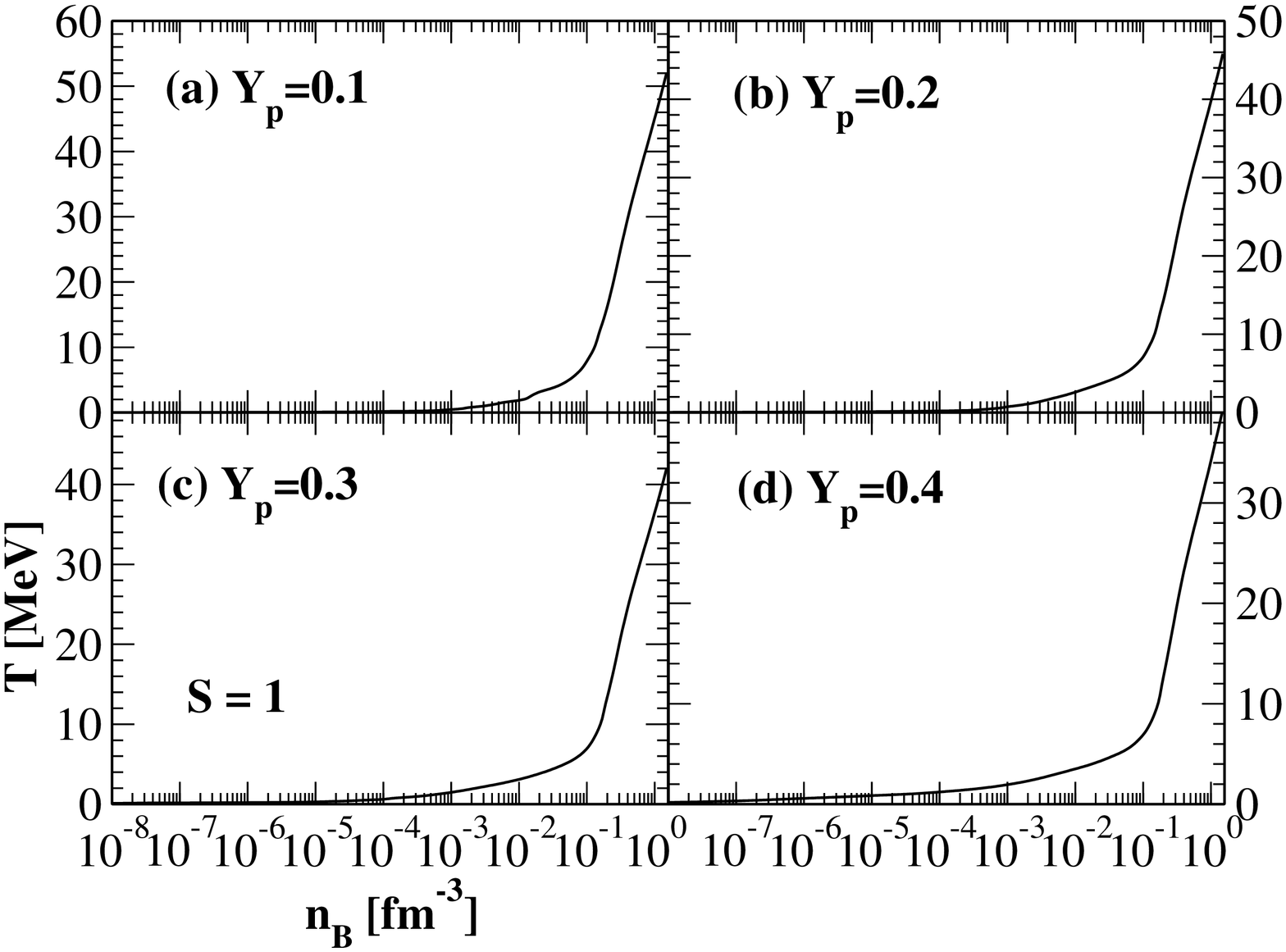}
 \includegraphics[height=8.5cm,angle=-90]{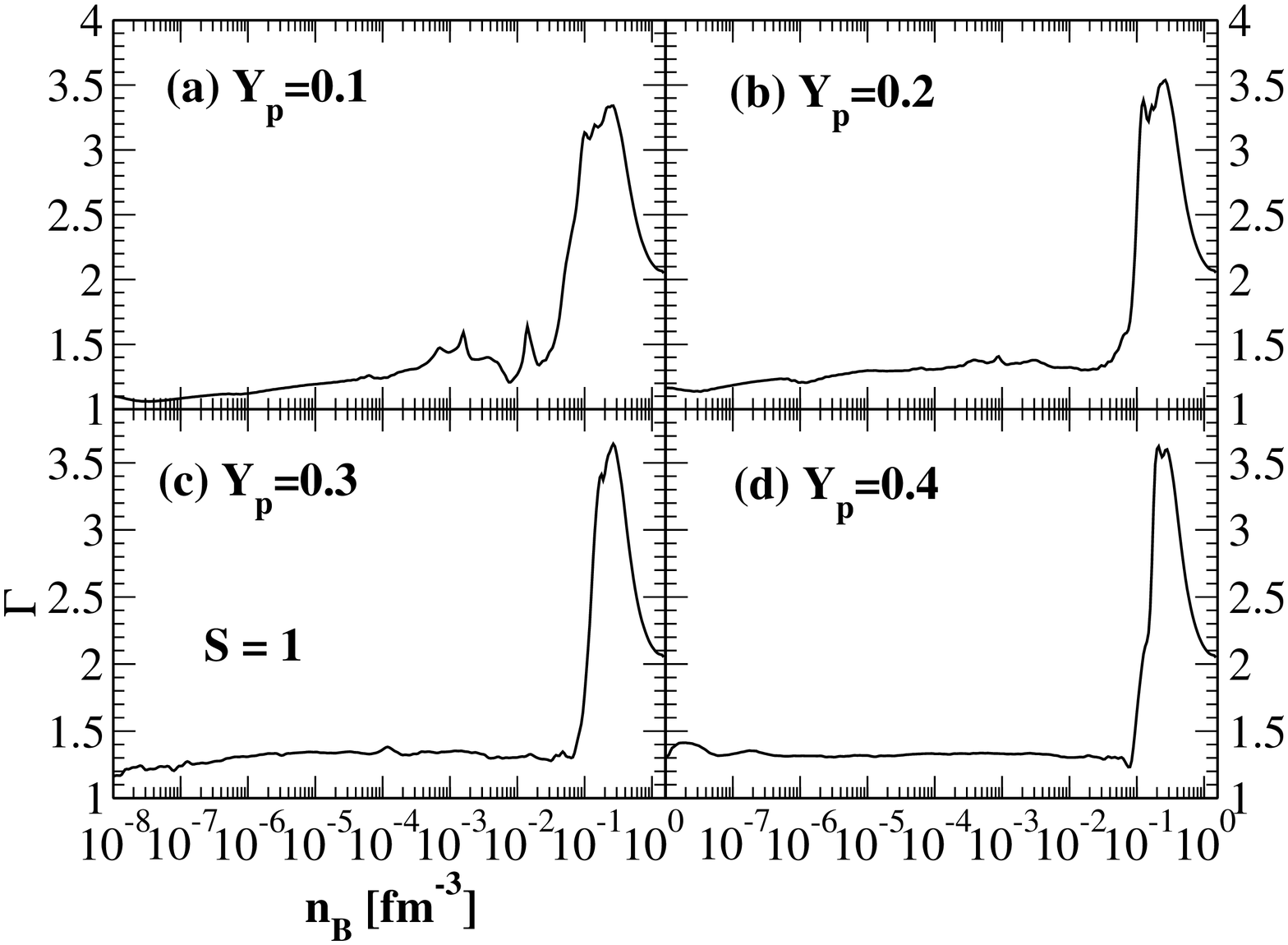}

\caption{(Color online) Left: temperature along adiabat $S$ = 1, for nuclear matter,
including baryon, electron, positron and photon, and proton
fraction $Y_P$ = 0.1 (a), 0.2 (b), 0.3 (c), and 0.4 (d). Right: adiabatic index for nuclear matter, including baryon,
electron, positron and photon, with constant entropy $S$ = 1, and
proton fraction $Y_P$ = 0.1 (a), 0.2 (b), 0.3 (c), and 0.4 (d).}\label{fig:adiabatic_gamma}
\end{figure}

%\end{widetext}

%\begin{widetext}

\begin{figure}[htbp]
 \centering
 \includegraphics[height=8.5cm,angle=-90]{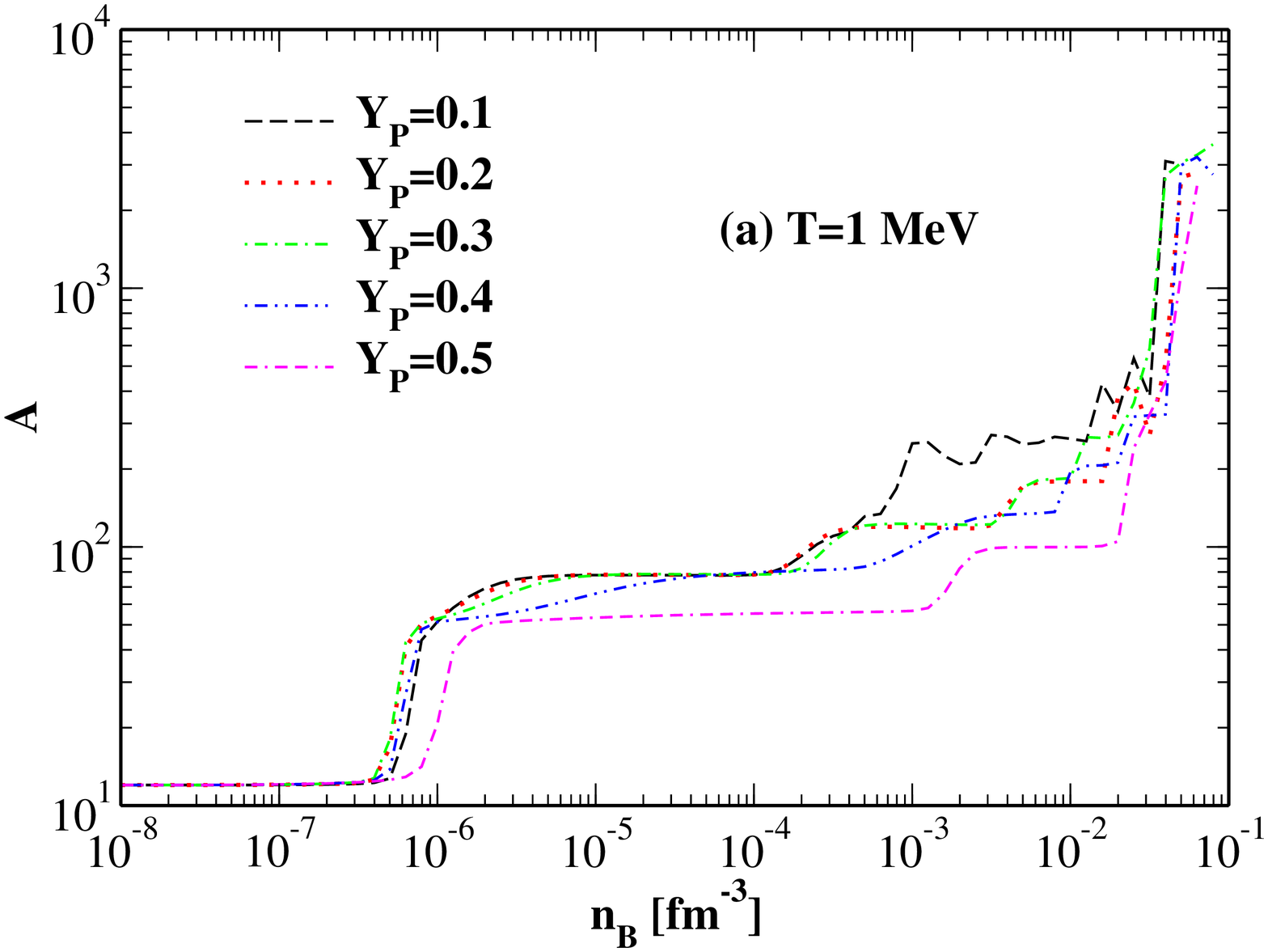}
 \includegraphics[height=8.5cm,angle=-90]{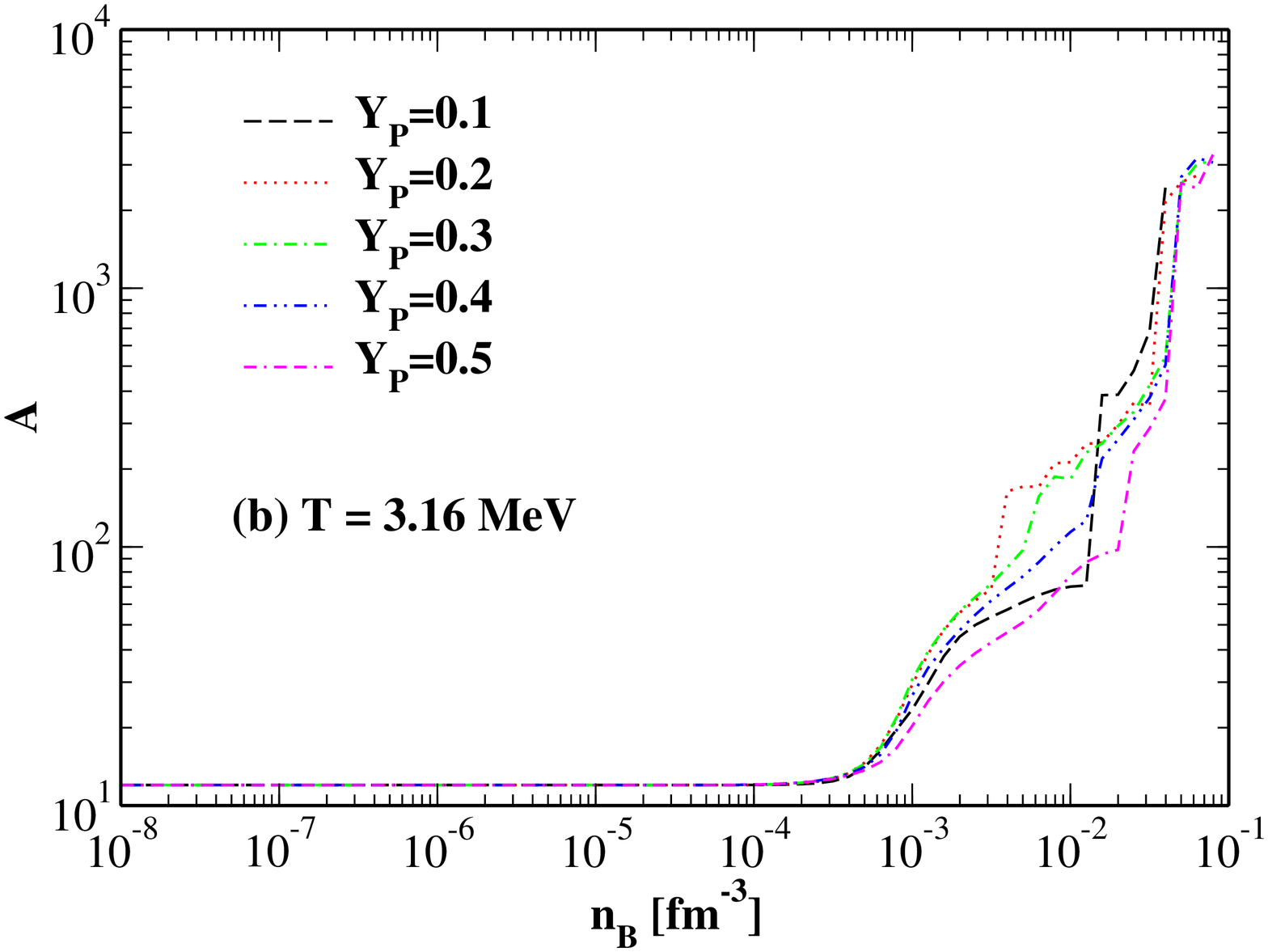}
 \includegraphics[height=8.5cm,angle=-90]{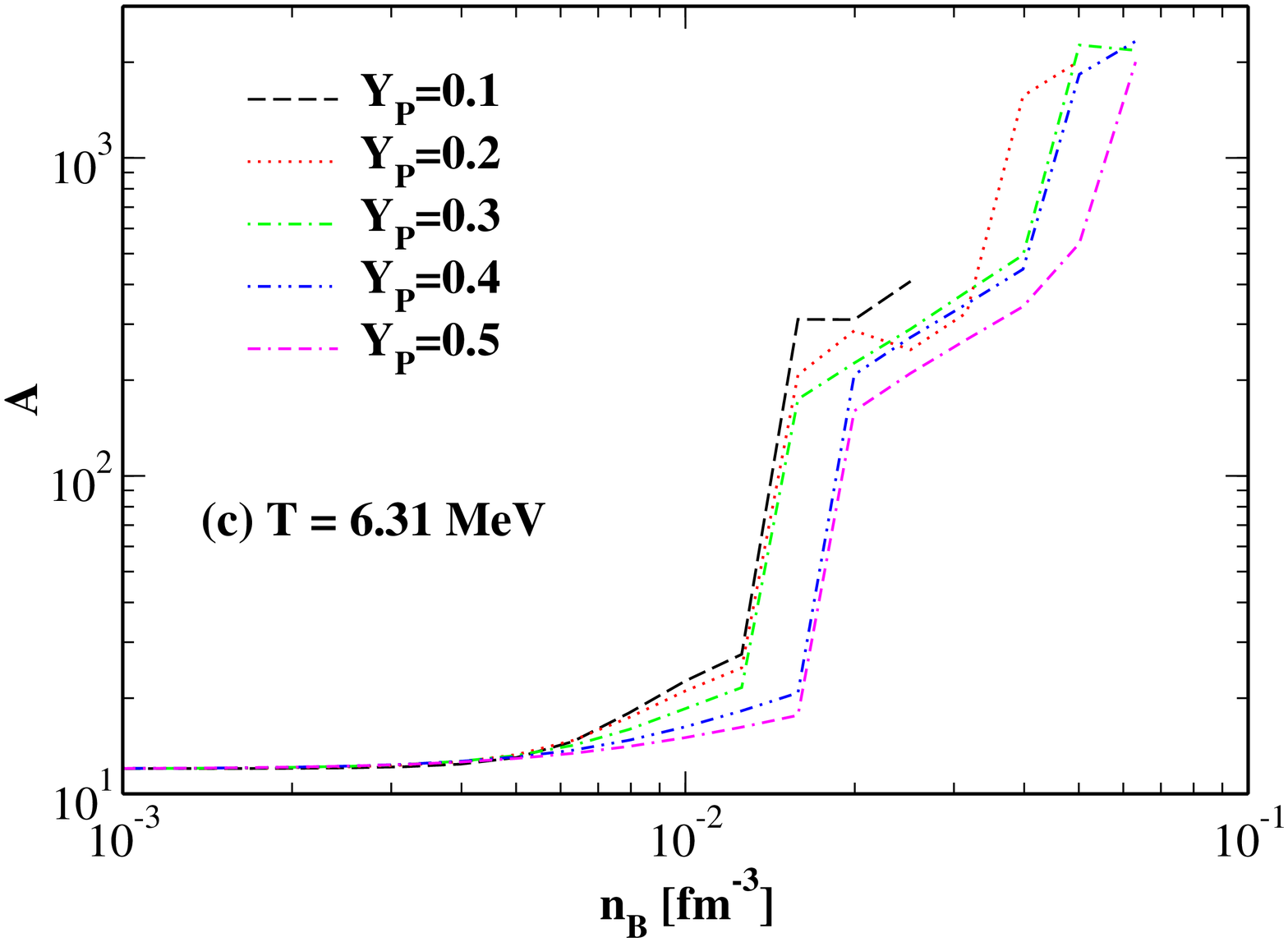}
 \includegraphics[height=8.5cm,angle=-90]{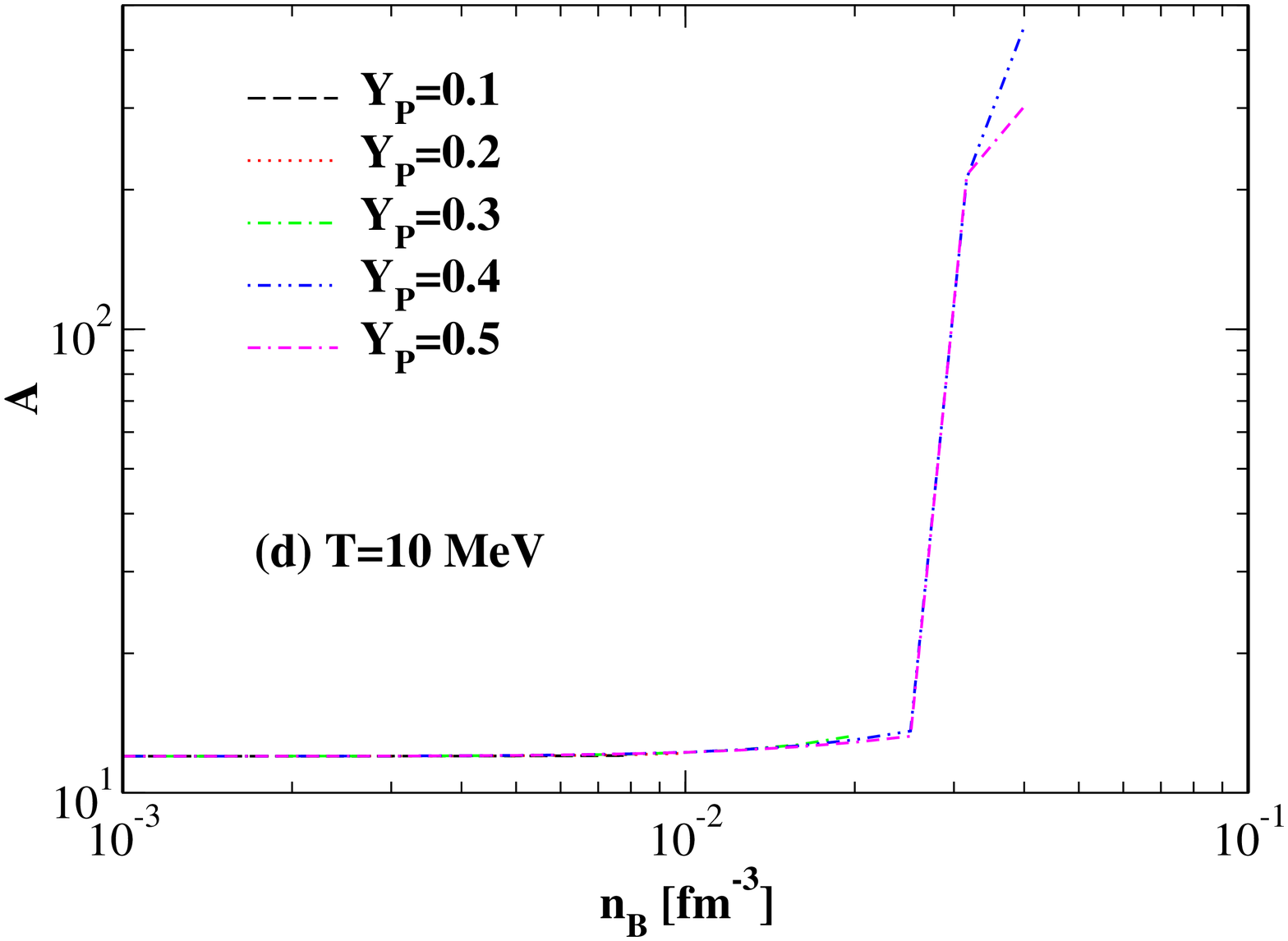}

\caption{(Color online) Average mass number of heavy nuclei at temperatures of T = 1 (a), 3.16 (b), 6.31 (c), and 10 (d) MeV. The proton fraction ranges from Yp = 0.1 to 0.5.}\label{fig:massnumber}
\end{figure}

%\end{widetext}

% proton number

%\begin{widetext}

\begin{figure}[htbp]
 \centering
 \includegraphics[height=8.5cm,angle=-90]{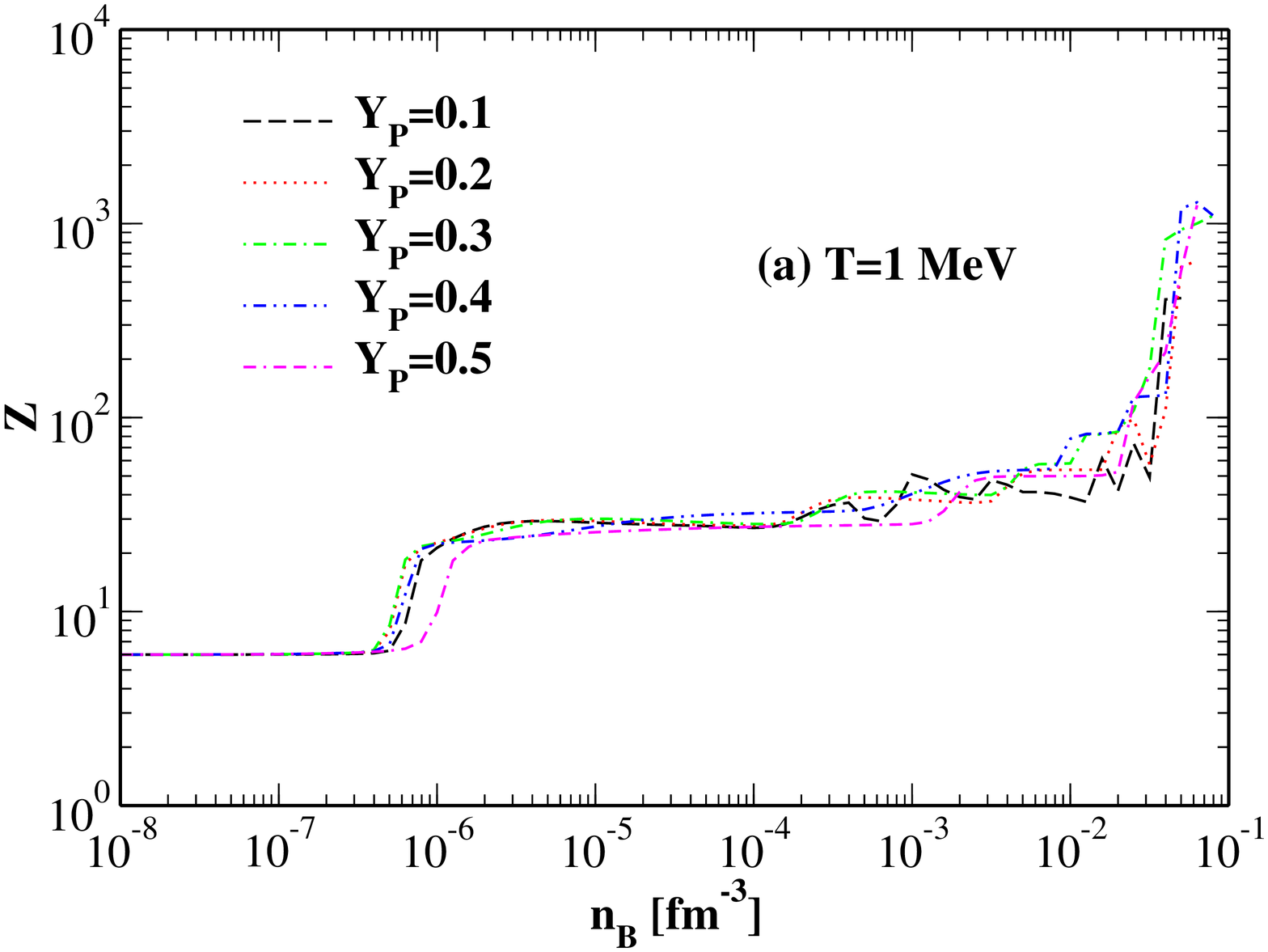}
 \includegraphics[height=8.5cm,angle=-90]{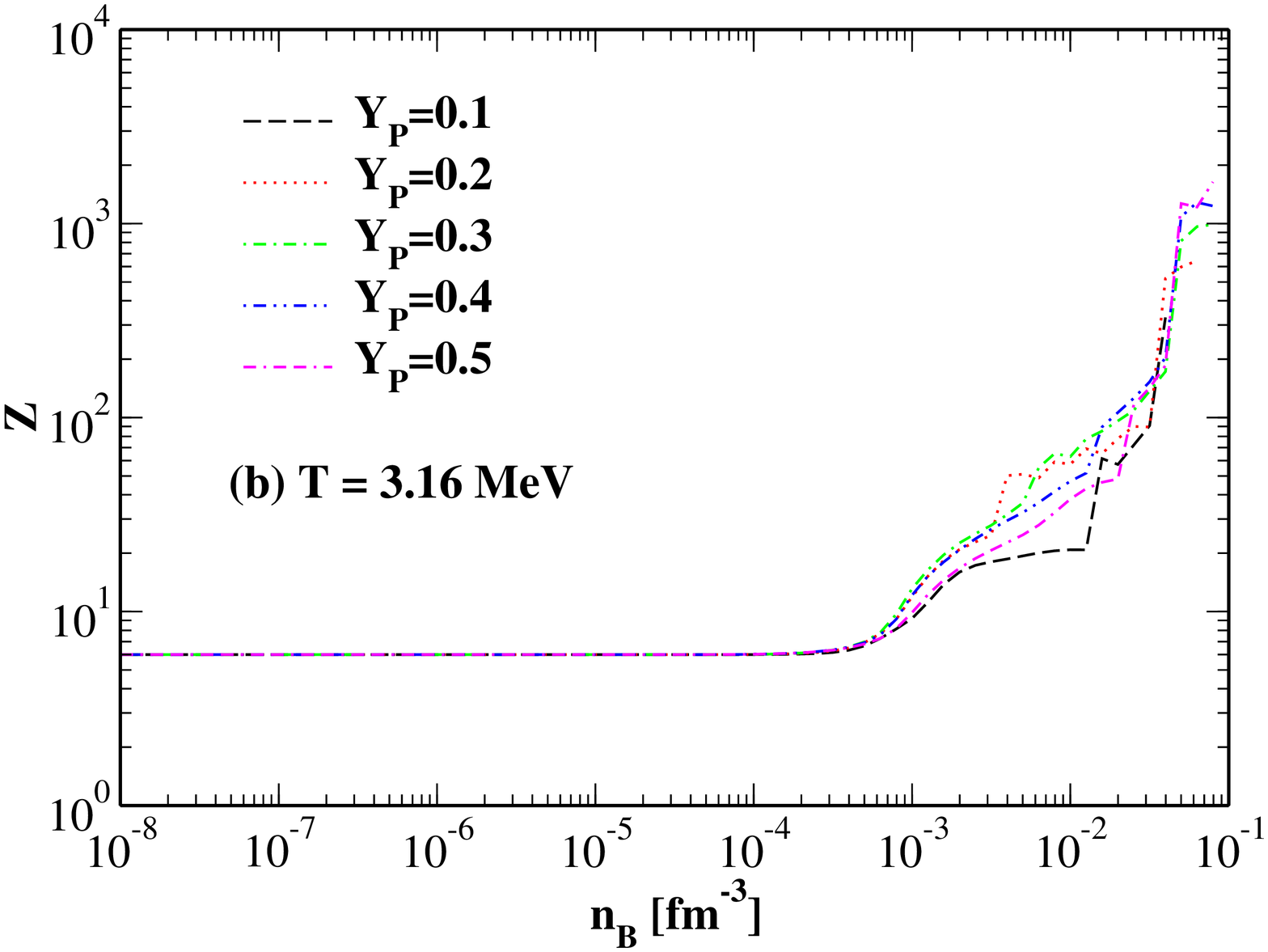}
 \includegraphics[height=8.5cm,angle=-90]{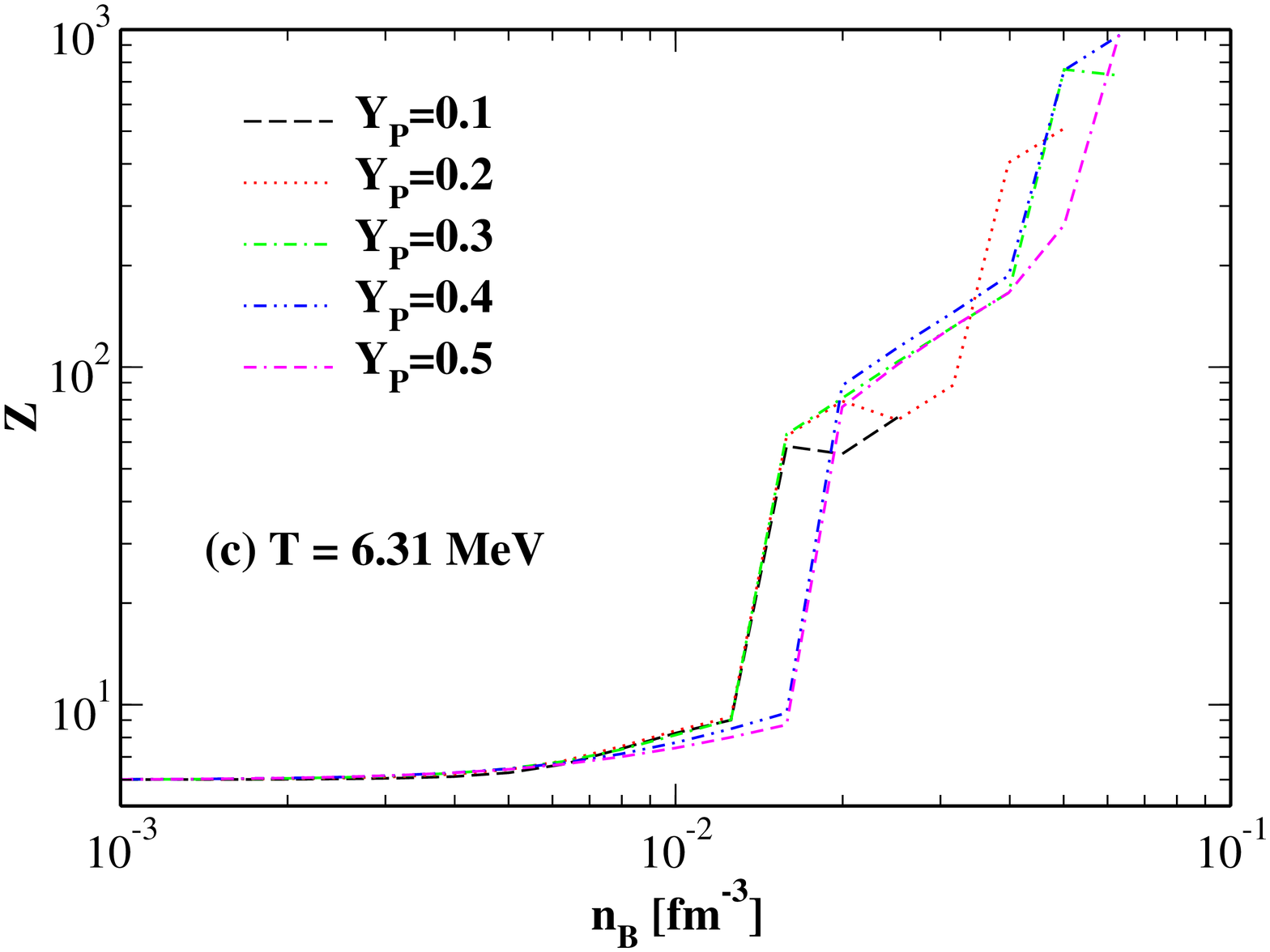}
 \includegraphics[height=8.5cm,angle=-90]{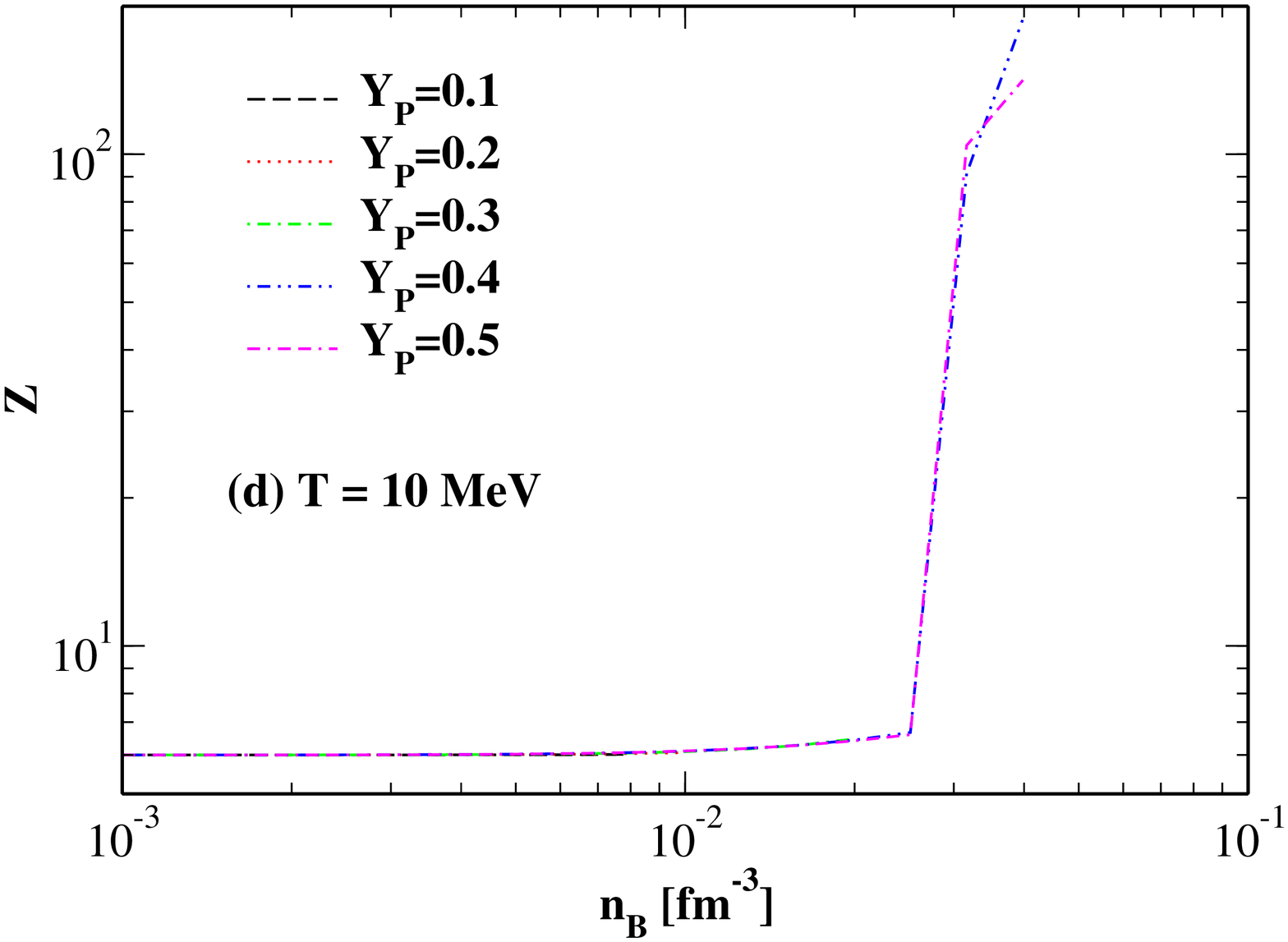}

\caption{(Color online) Average proton number of heavy nuclei at temperatures of T = 1 (a), 3.16 (b), 6.31 (c), and 10 (d) MeV. The proton fraction ranges from Yp = 0.1 to 0.5.}\label{fig:protonnumber}
\end{figure}

\end{widetext}

\subsection{Mass fractions of nucleons and different nuclei.}

% fractions: heavies, alpha, n, p; dominating heavies fraction

In Fig.~\ref{fig:fractiony05}, the mass fractions of free neutrons,
free protons, alpha particles and heavy nuclei are shown versus baryon density at four
different temperatures for a proton fraction $Y_p$ = 0.05. The
matter is extremely neutron rich, so the mass fraction of free
neutrons is always appreciable at any density.  Note that in the Hartree mean field
results there is only a single nucleus associated with each
Wigner-Seitz cell.  We define a nucleon level to be free when it has positive
energy.  The upper left panel is for $T$ = 1 MeV.  At very low
densities - less than a few times $10^{-8}$ fm$^{-3}$, the system is
basically  a free neutron and proton gas.  The alpha fraction rises
rapidly to 10\% around a density of $10^{-7}$ fm$^{-3}$.  Some of the free neutrons
are used to form alphas, so the fraction of free neutrons drops to 90\%.
The free neutron fraction drops further beyond a density of $10^{-6}$ fm$^{-3}$,
where heavy nuclei appear with a mass fraction of about 15\%.  The fraction of
heavy nuclei grows faster after the Virial gas - Hartree mean field
transition around 3$\cdot 10^{-4}$ fm$^{-3}$. More and more free
neutrons are bound to form heavier nuclei, as can be seen in
Fig.~\ref{fig:massnumber}.  The fraction of heavy nuclei reaches a maximum
value of 70\% before the final transition to uniform matter. The other
panels give the mass fractions of different species at higher
temperatures.  In these cases, the appearance of alpha particles and
heavy nuclei occur at higher densities and with smaller mass fractions compared to the $T$ = 1 MeV case. When $T$ = 6.31 MeV, there is only a narrow range of densities where alpha
particles have a mass fraction of a couple of percent and the amount of heavy nuclei is very small.

Figs.~\ref{fig:fractiony2} and \ref{fig:fractiony5} show the mass
fractions as were shown in Fig.~\ref{fig:fractiony05}, but
for matter with higher proton fractions $Y_p$ = 0.2 and 0.5,
respectively.  Alpha particles appear at similar densities as for $Y_p$ = 0.05, but with much larger mass fractions. For example, the alpha fraction can almost reach 100\% around $10^{-6}$ fm$^{-3}$ when $T$ = 1 MeV and $Y_P$ = 0.5.  Heavy nuclei also appear at similar densities as in $Y_p$ = 0.05, and with much larger mass fractions.  As the proton
fraction rises, heavy nuclei can survive to higher temperatures, even
though only at very high density. For example, when $T$ = 10 MeV
and $Y_P$ = 0.5, heavy nuclei have a large mass fraction for densities around a few
times 10$^{-2}$ fm$^{-3}$.

\subsection{Beta equilibrium at zero temperature.}

Matter in beta equilibrium at zero temperature is particularly
interesting and determines neutron star structure,
including the mass-radius relation and the neutron star crust. The $T$ = 0
EOS is included in our table \ref{tab:phasespace2}, which can be
used to find the beta equilibrium EOS by imposing the
constraint $\mu_n = \mu_p + \mu_e$.  The resulting energy per
baryon, pressure, and proton fraction are shown as function of
density in Fig.~\ref{fig:T0beta}.  The energy per baryon starts near
-8.5 MeV, a typical value for stable nuclei, at 10$^{-8}$ fm$^{-3}$,
becomes positive around 10$^{-3}$ fm$^{-3}$, and finally rises
rapidly after the transition to bulk uniform matter around 0.05
fm$^{-3}$.   The pressure vs density is polytropic like below
10$^{-4}$ fm$^{-3}$, and then it develops small plateaus.  After the 
transition to uniform matter, the pressure rises more rapidly
which indicates a stiffening of the EOS.  At 10$^{-8}$ fm$^{-3}$ matter
is made up of stable nuclei with similar numbers of neutrons and
protons.  That's why the proton fraction at 10$^{-8}$ fm$^{-3}$ is
as high as 0.45.  The proton fraction drops as the density increases, which
means matter becomes more and more neutron rich due to the large electron Fermi energy. The proton fraction reaches a minimum value of 0.007 around 0.025 fm$^{-3}$.  However, it increases again due to the large symmetry energy at high density for the NL3 \cite{NL3} interaction, which favors more nearly equal neutron and proton fractions.

\subsection{Neutron star structure}

The mass and radius of a neutron star are obtained by solving the 
Tolman-Oppenheimer-Volkoff (TOV) equations \cite{oppenheimer39,tolman39}.
%\beqn\label{O-V}
 %    &&\frac{dp(r)}{dr} =
  %  - \frac{[p(r) + \epsilon(r)][M(r) + 4\pi r^3p(r)]}{r(r - 2M(r))}\, ,\\
%     &&M(r) = 4\pi\int_0^r\epsilon(r^\prime)r^{\prime2}dr^\prime\, ,
%\eeqn 
%where $r$ denotes the radial coordinate relative to the
%center of the star. $p(r)$ and $\epsilon(r)$ are the pressure and
%energy density at a radial point $r$ in the star respectively,
%which are obtained in previous section. $M(r)$ represents the mass
%of the sphere contained within a radius $r$. The radius, $R$, of
%the star is defined as that radius at which the pressure is zero.
%The total mass, $M(R)$, of the star is subsequently defined as the
%mass contained within a sphere of radius $R$. 

By using the EOS in beta equilibrium at zero temperature from the previous section in the core and using Baym-Pethick-Sutherland EOS in the crust \cite{BPS}, we
obtain the neutron star mass-radius and mass-central density
relations, which are plotted in Fig.~\ref{fig:m-r}. The neutron star's maximum gravitational mass is about 2.77 solar mass with a radius of 13.3
km. The corresponding central density is about 10$^{15}$ g/cm$^3$.
The large value for the maximum neutron star mass is not too
surprising since NL3 provides a very stiff EOS as we have
emphasized several times.

\subsection{Comparisons with Lattimer-Swesty's EOS and H. Shen \etal's EOS}

It is instructive to compare our EOS with the Lattimer-Swesty equation of state, that uses a simple liquid drop model with a non-relativistic Skyrme interaction, and the H. Shen \etal     equation of state, that uses the Thomas Fermi and variational
approximations to a relativistic mean field model. The L-S EOS quoted in this work corresponds to the one with incompressibility coefficient $K$=180 MeV. 
In Fig.~\ref{fig:p_comp}, the pressure for matter at
$T$ = 1, or 6.31 MeV and $Y_P$ = 0.05 or 0.4 is shown for our
EOS, Lattimer-Swesty's (L-S) and H. Shen \etal's (S-S) EOSs. Here
the pressure includes contributions from electrons, positrons and
photons. Below nuclear saturation density $\sim$ 0.16 fm$^{-3}$,
ours and S-S EOS agree very well. The L-S EOS gives a slightly lower
pressure at densities above 10$^{-3}$ fm$^{-3}$ when $T$ = 1 MeV and $Y_P$ =
0.05.  Above saturation density, our EOS gives the largest pressure.

Fig.~\ref{fig:xh_comp} compares the mass fraction of heavy nuclei from
our EOS, L-S EOS and S-S EOS, for matter at $T$ = 1 or 6.31 MeV,
and $Y_P$ = 0.05 or 0.4.  Although the pressure agrees well among the three EOSs, the mass fraction of heavy nuclei and alpha particles can be different and this contributes to the differences in entropy shown in Fig.~\ref{fig:s_comp}.  In addition our EOS may have a higher entropy in Fig.\ref{fig:s_comp} (b) because we may have smaller nuclei for $n$ less than $10^{-3}$ fm$^{-3}$ compared to S-S and because we can have a distribution of several heavy nuclei, while the L-S and S-S EOSs use only a single average species.   The average $A$ and $Z$ of heavy nuclei is shown in Fig.~\ref{fig:AZ}.  This figure clearly shows the effects of shell structure, that is included in our EOS but is neglected in both L-S and S-S EOSs.   As a result, $A$ and $Z$ for our EOS have a serious of steps while $A$ and $Z$ for L-S and S-S EOSs increase smoothly with density. 

For $T$ = 1 MeV and $Y_P$ = 0.05, the three EOSs agree very well on
entropy, except that at low densities the L-S EOS gives a slightly smaller
value. It is comprehensible since the matter is dominated by
free neutrons until very high density when heavy nuclei dominate
(upper left panel of Fig.~\ref{fig:xh_comp}). But heavy nuclei
have low entropy at low temperatures so the difference in entropy between the EOSs is small.  For $T$ = 1 MeV and $Y_P$ = 0.4, the differences in entropy between the three EOSs are larger.  This is partially due to the smaller scale used. It is also because the
different EOS predict different mass fractions of heavy nuclei, whose formation decreases the entropy. The L-S EOS has 80\% heavy nuclei at 10$^{-6}$ fm$^{-3}$ (upper right panel in
Fig.~\ref{fig:xh_comp}) compared to 20\% in our EOS and the S-S EOS,
therefore the L-S EOS has the lowest entropy.  At the higher temperature $T$ =
6.31 MeV, the three EOSs agree well on entropy for different values of
$Y_P$, because free nucleons dominate at most densities (see lower panels in
Fig.~\ref{fig:xh_comp}).

The differences in abundances and average $A$ and $Z$ between EOSs may arise because of differences in symmetry energy and approximations made.  For example L-S uses a very simple liquid drop model while the S-S EOS uses variational and Thomas Fermi approximations that are avoided in our EOS.  These differences could be important for neutrino interactions and should be studied further.

\begin{widetext}

\begin{figure}[htbp]
 \centering
 \includegraphics[height=13cm,angle=-90]{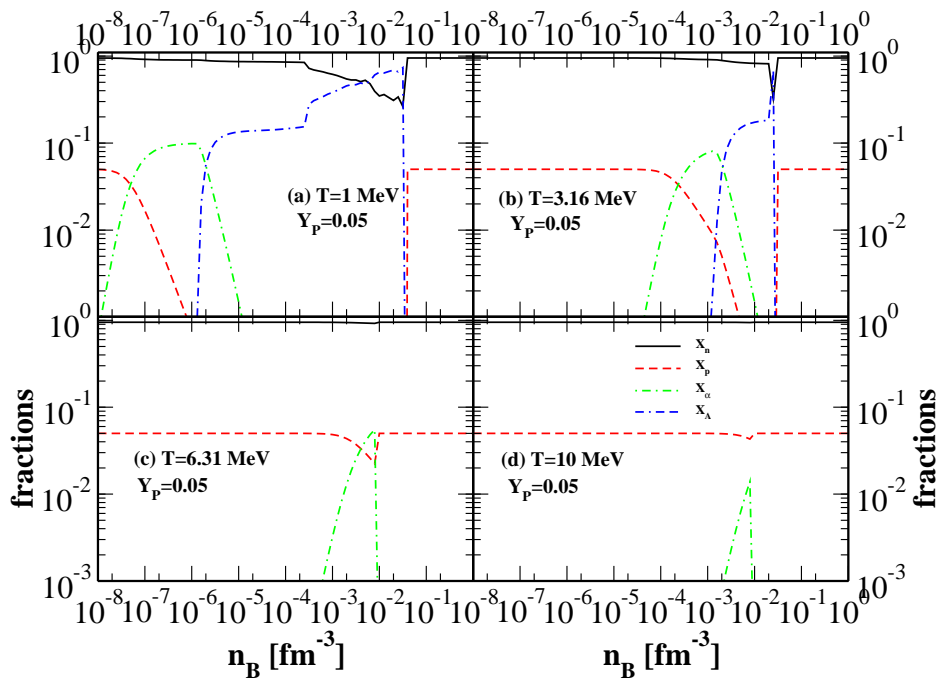}

\caption{(Color online) Mass fractions of free neutrons, free protons, alpha
particles and heavy nuclei versus baryon density at four different
temperatures T = 1 (a), 3.16 (b), 6.31 (c), and 10 (d) MeV, and fixed proton fraction $Y_p$ =
0.05.}\label{fig:fractiony05}
\end{figure}

\begin{figure}[htbp]
 \centering
 \includegraphics[height=13cm,angle=-90]{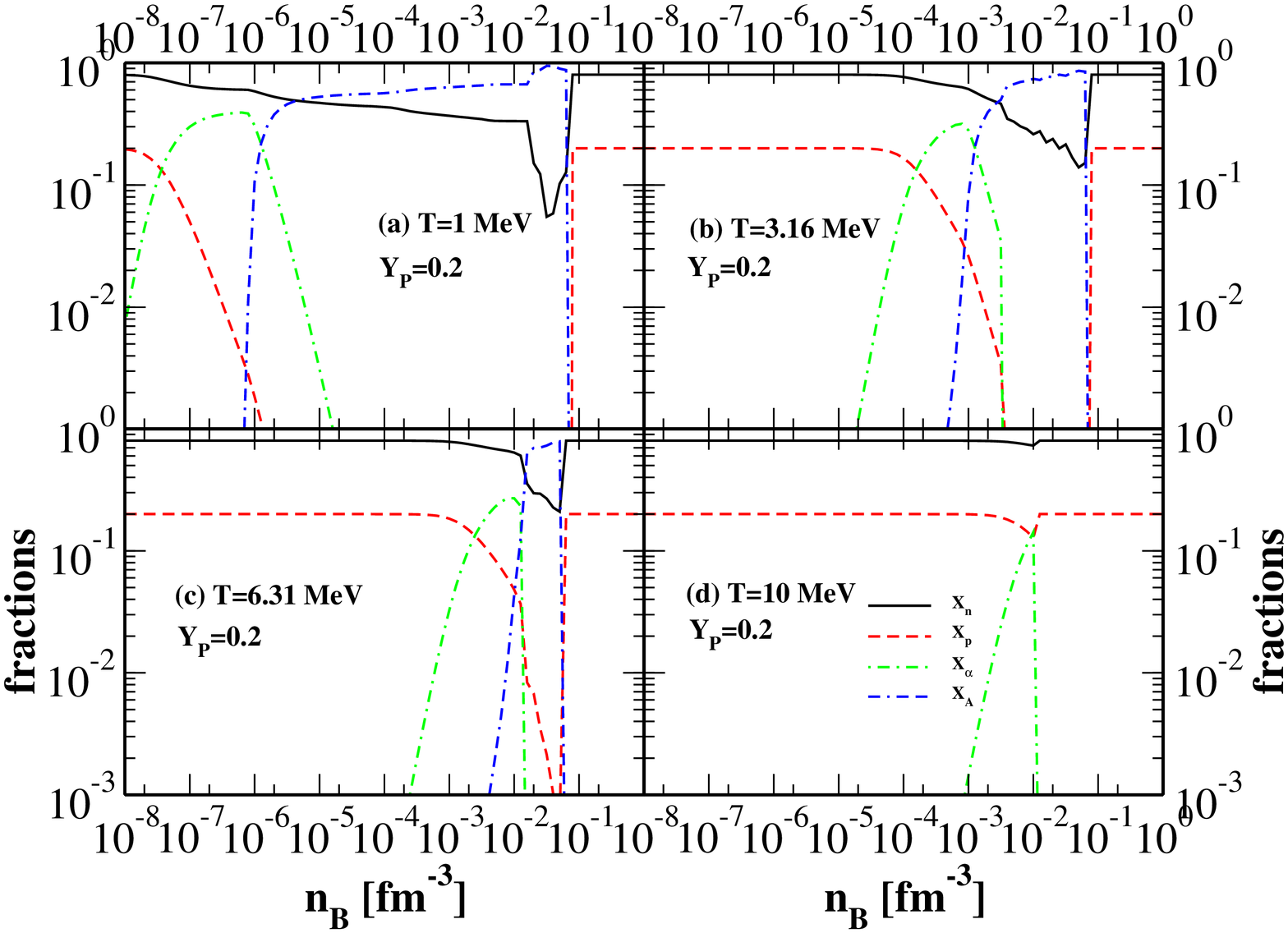}

\caption{(Color online) Mass fractions of free neutrons, free protons, alpha
particles and heavy nuclei versus baryon density at four different
temperatures T = 1 (a), 3.16 (b), 6.31 (c), and 10 (d) MeV, and fixed proton fraction $Y_p$ =
0.2.}\label{fig:fractiony2}
\end{figure}

\begin{figure}[htbp]
 \centering
 \includegraphics[height=13cm,angle=-90]{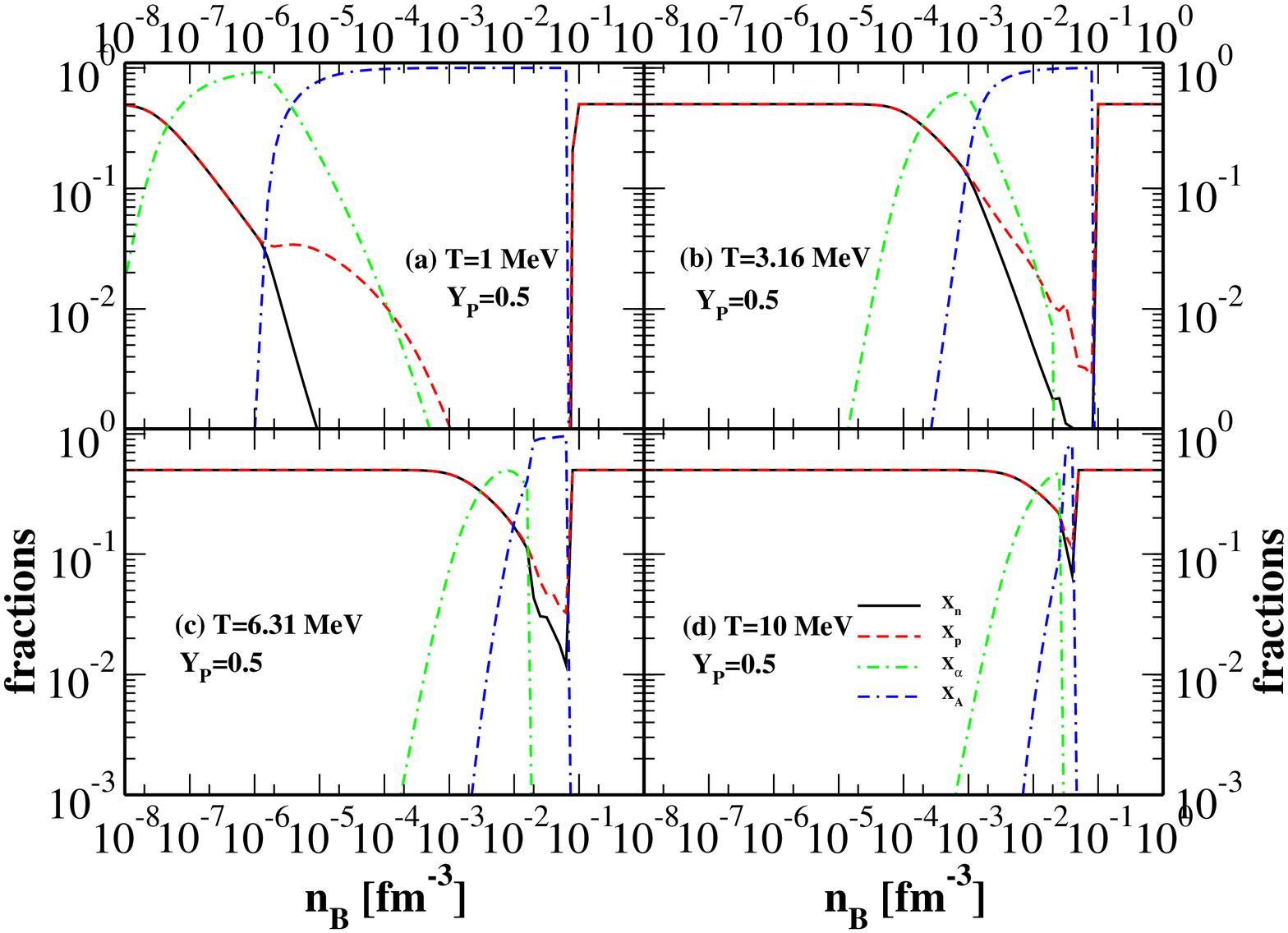}

\caption{(Color online) Mass fractions of free neutrons, free protons, alpha
particles and heavy nuclei versus baryon density at four different
temperatures T = 1 (a), 3.16 (b), 6.31 (c), and 10 (d) MeV, and fixed proton fraction $Y_p$ =
0.5.}\label{fig:fractiony5}
\end{figure}

%\end{widetext}

%\begin{widetext}

\begin{figure}[htbp]
 \centering
 \includegraphics[height=8.5cm,angle=-90]{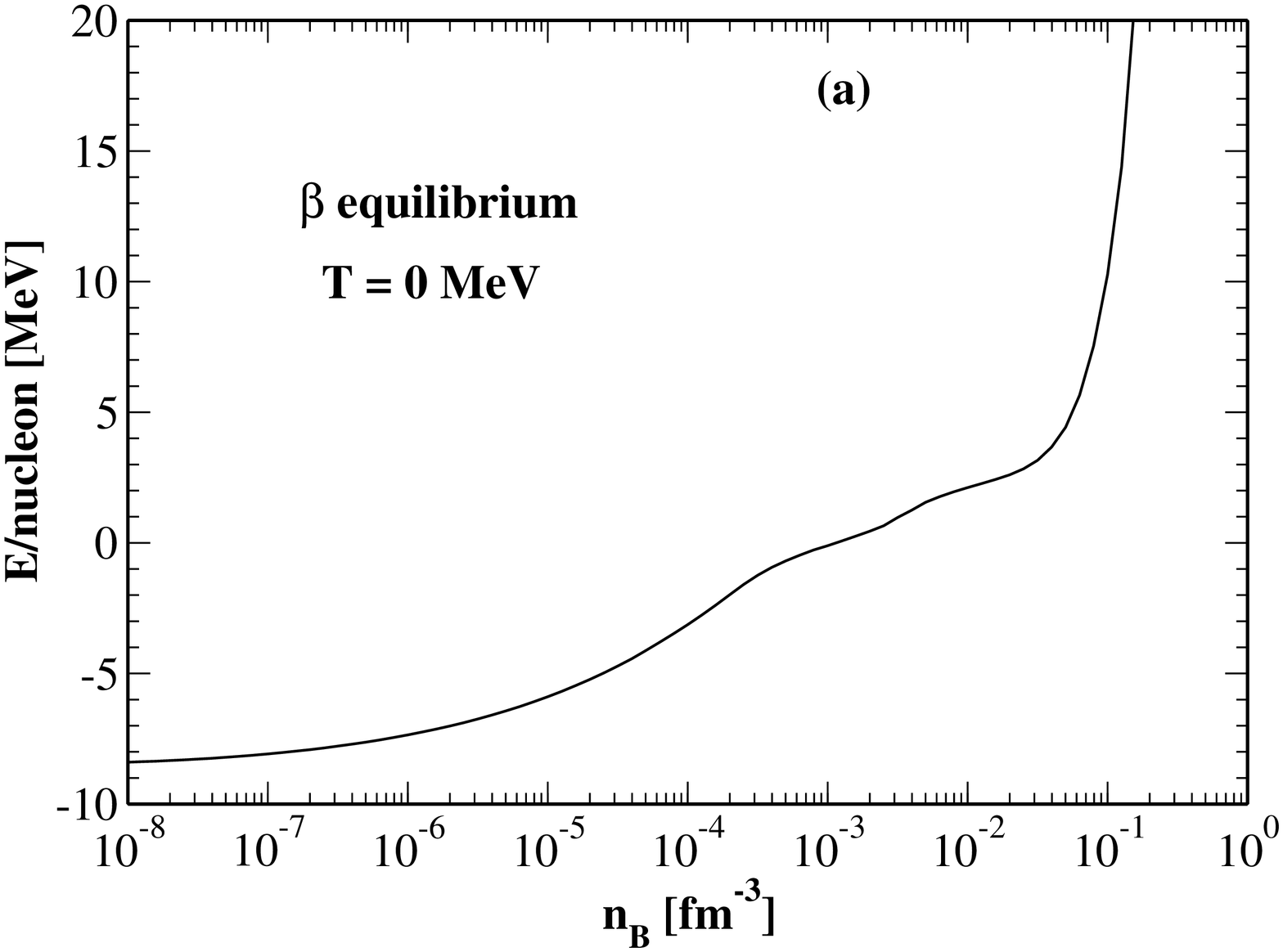}
 \includegraphics[height=8.5cm,angle=-90]{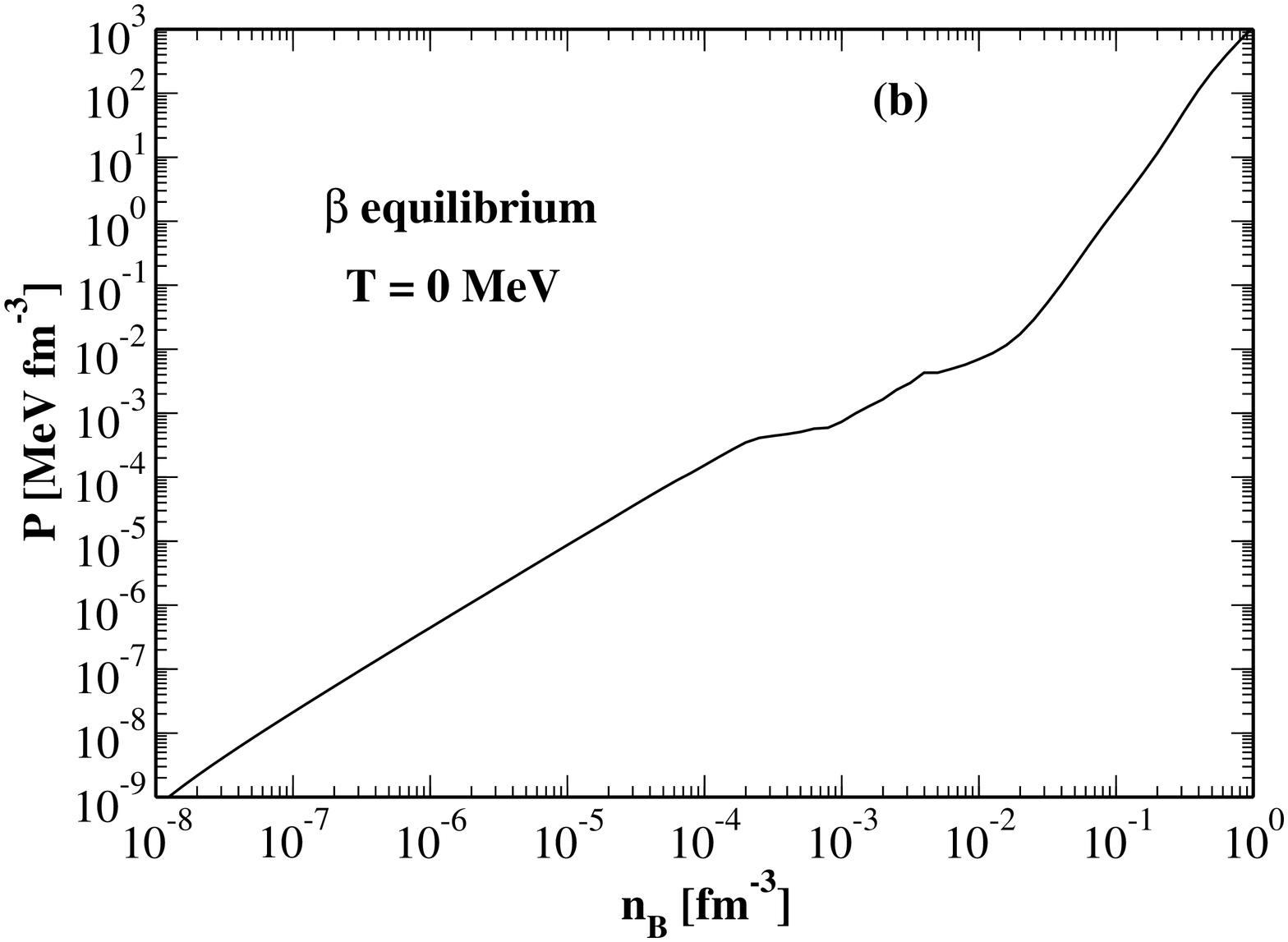}
 \includegraphics[height=8.5cm,angle=-90]{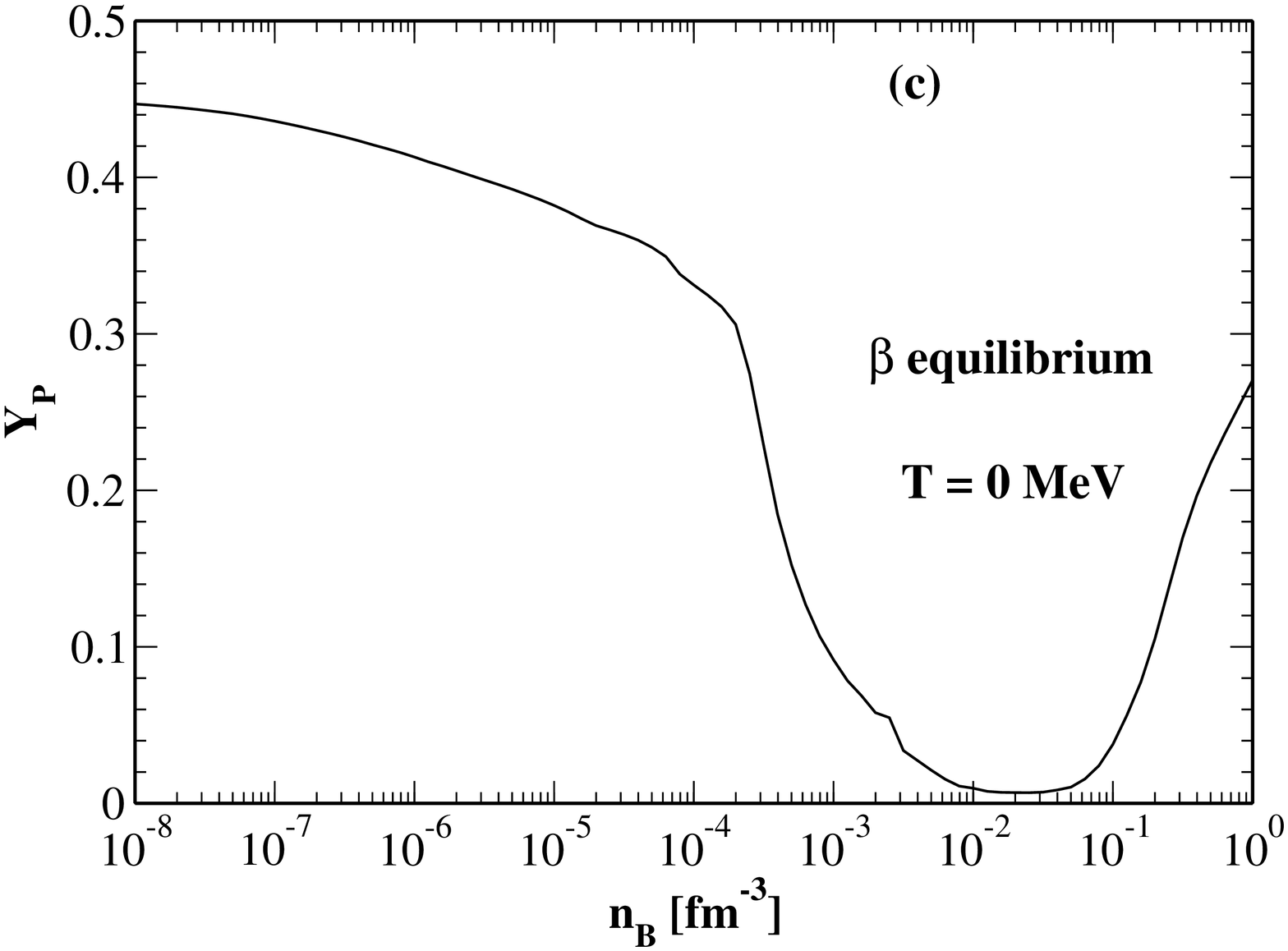}

\caption{Energy (a), pressure (b) and proton fraction (c) versus density for
zero temperature nuclear matter in chemical equilibrium.
}\label{fig:T0beta}
\end{figure}

%\end{widetext}

%\begin{widetext}

\begin{figure}[htbp]
 \centering
 \includegraphics[height=13cm,angle=-90]{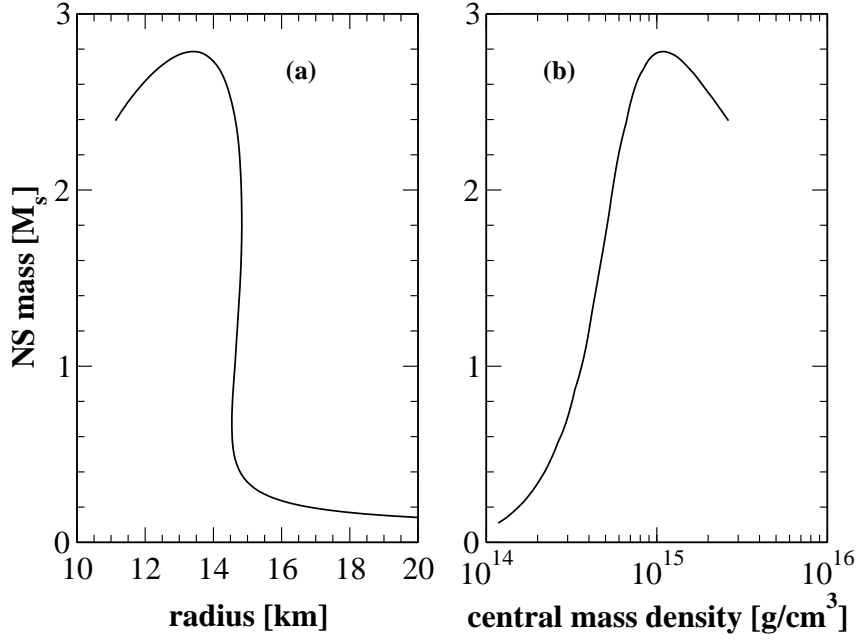}

\caption{Mass vs Radius (a) and mass vs central density
(b) for neutron star from zero temperature beta
equilibrium nuclear matter as shown in
Fig.~\ref{fig:T0beta}.}\label{fig:m-r}
\end{figure}

%\end{widetext}

%\begin{widetext}

\begin{figure}[htbp]
 \centering
 \includegraphics[height=13cm,angle=-90]{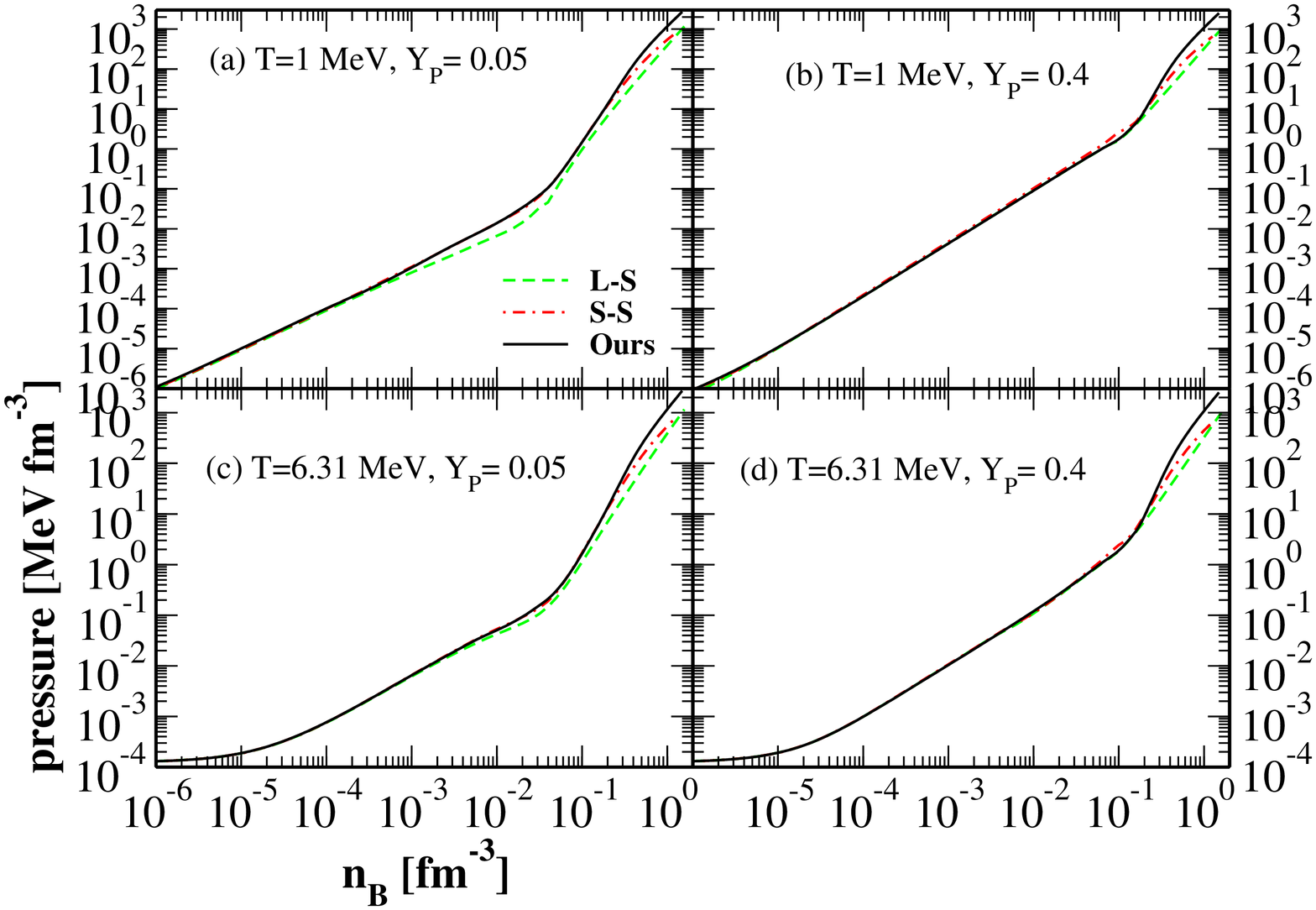}

\caption{(Color online) Pressure versus baryon density from our EOS,
Lattimer-Swesty's (L-S) and H. Shen \etal's (S-S)
EOS, with T = 1 MeV,  $Y_p$ = 0.05 (a), T = 1 MeV,  $Y_p$ = 0.4 (b), T = 6.31 MeV,  $Y_p$ = 0.05 (c), and T = 6.31 MeV,  $Y_p$ = 0.4 (d).}\label{fig:p_comp}
\end{figure}

\begin{figure}[htbp]
 \centering
 \includegraphics[height=13cm,angle=-90]{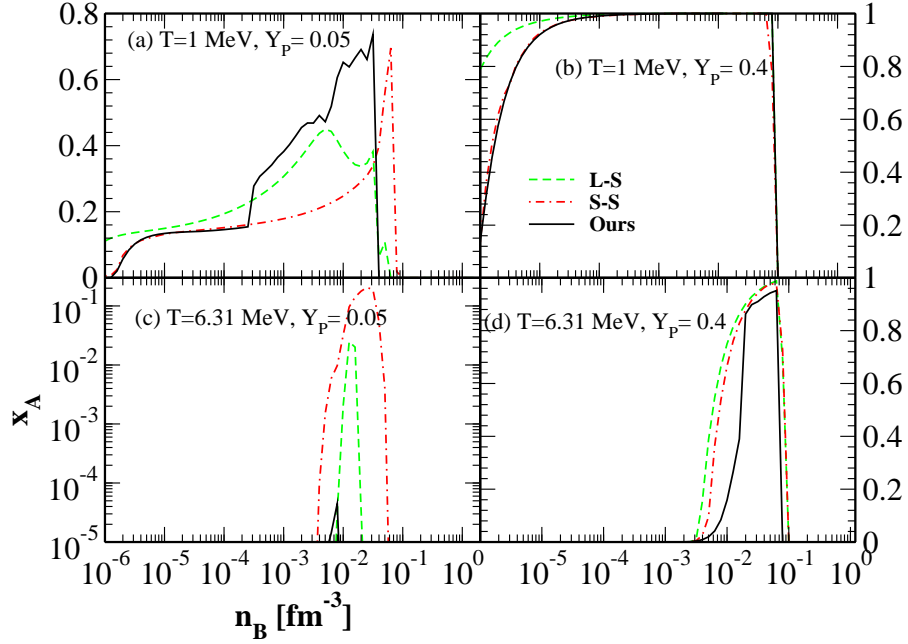}

\caption{(Color online) mass fraction of heavy nuclei versus baryon density from
our EOS, Lattimer-Swesty's and H. Shen \etal's
EOS, with T = 1 MeV,  $Y_p$ = 0.05 (a), T = 1 MeV,  $Y_p$ = 0.4 (b), T = 6.31 MeV,  $Y_p$ = 0.05 (c), and T = 6.31 MeV,  $Y_p$ = 0.4 (d).}\label{fig:xh_comp}
\end{figure}

\begin{figure}[htbp]
 \centering
 \includegraphics[height=13cm,angle=-90]{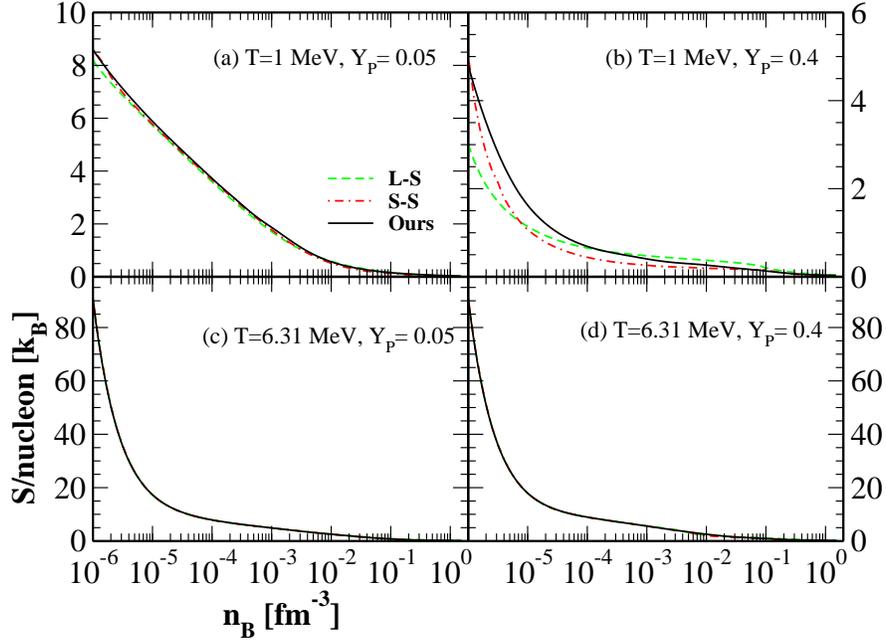}

\caption{(Color online) Entropy per baryon versus baryon density from our EOS,
Lattimer-Swesty's and H. Shen \etal's EOS, with T = 1 MeV,  $Y_p$ = 0.05 (a), T = 1 MeV,  $Y_p$ = 0.4 (b), T = 6.31 MeV,  $Y_p$ = 0.05 (c), and T = 6.31 MeV,  $Y_p$ = 0.4 (d).}\label{fig:s_comp}
\end{figure}

\begin{figure}[htbp]
 \centering
 \includegraphics[height=13cm,angle=-90]{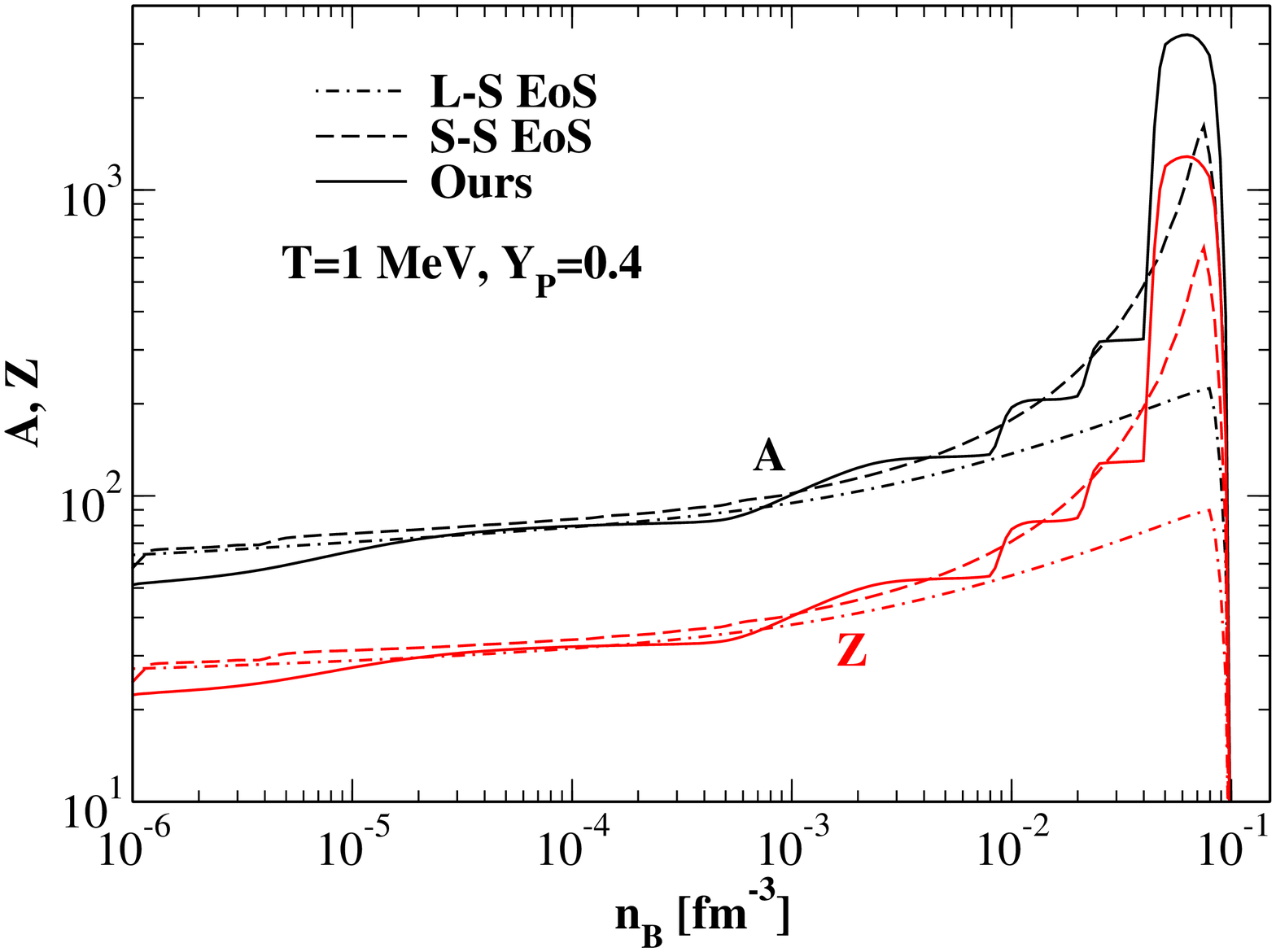}

\caption{(Color online) Average mass number $A$ (black upper curves) and atomic number $Z$ (red lower curves) of heavy nuclei for our EOS (solid) , Lattimer-Swesty's (dot-dashed) and H. Shen \etal's
EOS (dashed) for T = 1 MeV,  $Y_p$ = 0.4.}\label{fig:AZ}
\end{figure}

\end{widetext}

\subsection{Adabatic compression tests}

One important test of thermodynamic consistency for our EOS table is to check that entropy is conserved as matter undergoes adiabatic compression.   This is very closely related to the conservation of energy in the first law of thermodynamics.   Let us consider the following adiabatic compression procudure. 

\begin{enumerate}
\item Start from an energy per baryon $E(n(t_0),T(t_0))$, a density $n(t_0),$ and a temperature $T(t_0)$.  Here $t$ stands for a sequence of times starting at $t_0$ . The pressure $P$ is then given by the EOS.  
\item Now compress the matter by increasing the density by an amount $dn$ so that $n(t_1)=n(t_0)+dn$ with $t_1=t_0+1$.
\item Next ensure that the first law of thermodynamics is satisfied by using a backward difference equation and requiring 
\beq
E(n(t_1),T(t_1)) = E(n(t_0),T(t_0)) + P(n(t_1),T(t_1)) dn
\eeq
by solving for the new pressure $P(n(t_1),T(t_1))$ and temperature $T(t_1)$.	
\item Finally the updated entropy $S(n(t_1),T(t_1))$ follows from our EOS at the new density $n(t_1)$ and temperature $T(t_1)$. 

\end{enumerate}       

\begin{widetext}

\begin{figure}[htbp]
 \centering
 \includegraphics[height=13cm,angle=-90]{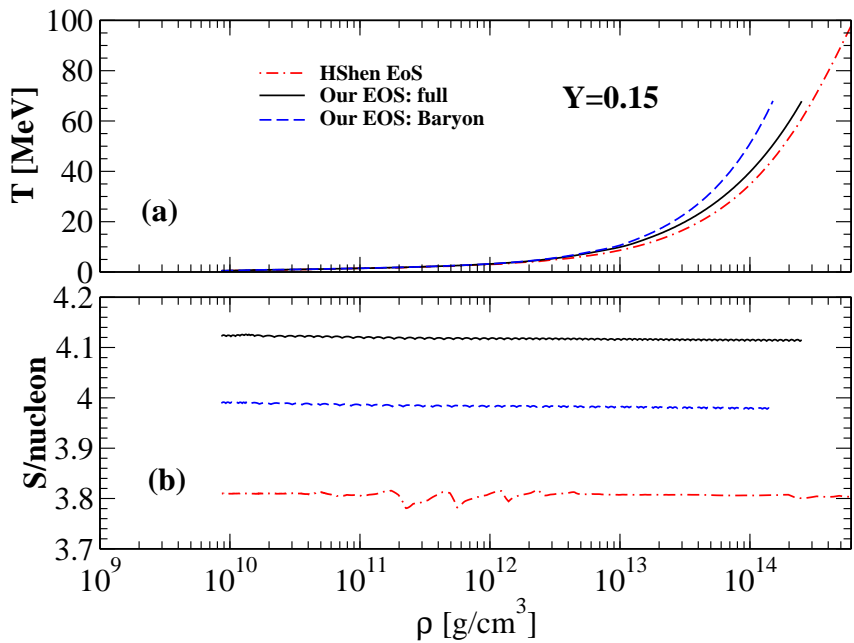}

\caption{(Color online) Temperature (a) and entropy (b) versus density in the adiabatic compression for nuclear matter with fixed proton fraction 0.15.}\label{fig:compressiony0.15}
\end{figure}

\end{widetext}

The entropy should be conserved in this adiabatic compression so that $S$ is independent of $t$.  In Fig.~\ref{fig:compressiony0.15}, the temperature and entropy versus density during  adiabatic compression are shown for nuclear matter with fixed proton fraction $Y_P =$ 0.15.
The initial density is 5.16738 $\times$ 10$^{-6}$ fm$^{-3}$ and initial temperature is 0.5 MeV. The solid curve is test result for our EOS table with lepton and photon contributions included in addition to contributions from baryons. The dashed curve is a test result for our EOS table with baryons alone.  The variation in entropy for both cases is less than 1\%. For a comparison the test result for H. Shen EOS is also shown (as dot-dashed curve) for the similar initial condition. Note that it is important to use an accurate interpolation scheme with the EOS table to ensure that the first law is satisfied so that entropy is conserved. The adiabatic compression test result for H. Shen EOS was obtained using the routine developed in Ref.~\cite{Ott09}.

\section{\label{summary}Summary and Outlook}

In this paper we generate a new complete equation of
state for use in supernova and neutron star merger simulations.
We use a density dependent RMF model for nuclear
matter at intermediate and high density with a spherical
Wigner-Seitz approximation for nonuniform matter, which
incorporates nuclear shell effects \cite{SHT10a}. For nuclear matter at
low density \cite{SHT10b}, we use a Virial expansion for a nonideal gas consisting
of neutrons, protons, alpha particles and thousands of heavy
nuclei from the finite range droplet model (FRDM) mass table \cite{FRDM}.  We include second order virial
corrections for light elements $A \leq$ 4, nuclear partition
functions for heavy nuclei, and Coulomb corrections between
electrons and heavy nuclei.  As the density decreases,
the mean field results match smoothly to the Virial gas.  As the density goes down, the Virial expansion reduces to nuclear statistical equilibrium and the Virial expansion is exact in the low density limit.  The computed EOS covers 180,000 grid points in the temperature range $T$ = 0 $\sim$ 80 MeV, the density range $n_B$ = 10$^{-8}$ $\sim$ 1.6 fm$^{-3}$, and the proton fraction range $Y_P$ = 0 $\sim$ 0.56.

We use a hybrid interpolation scheme to generate a full EOS table on a fine grid that is
thermodynamic consistent.   This ensures that the first law of thermodynamics is satisfied and that entropy is conserved during adiabatic compression.   Our EOS is an improvement over the existing  Lattimer-Swesty \cite{LS} and H. Shen \etal~ \cite{Shen98a,Shen98}, equations of state because our EOS includes thousands of heavy nuclei and is exact in the low density limit.

We discuss the thermodynamic properties of our EOS in detail.
The free energy per baryon, pressure, entropy per baryon and
chemical potentials for neutrons, protons, and electron neutrinos are shown for matter at high and low temperatures and for large and small proton fractions.  We also perform adiabatic compression tests for our EOS table and show that it preserves entropy to better than 1\%.  

Finally, we show some comparisons between our EOS and two
existing EOS tables. At high density, our RMF model reduces to
the normal NL3 set. It produces a stiffer EOS at high density than the
S-S EOS, which is already stiffer than the L-S EOS. One can find this
difference, for example, in Fig.~\ref{fig:p_comp} for the pressure at
high density.  As a result, neutron stars obtained from our EOS have a
large maximum mass of around 2.77 solar mass and radius of 13.3 km.
At low density, our model is essentially the virial expansion of a 
nonideal gas, which includes many nuclei.  Our model usually has a broad
distribution of nuclei as shown in
our previous paper \cite{SHT10b}.  This spread in
the distribution of nuclei makes the composition of nuclear matter in
our EOS different from that in the L-S and S-S EOS tables.  This composition difference also
introduces some differences in the entropy for nuclear matter at
low temperature and large proton fraction, see for example the upper
right panel in Fig.~\ref{fig:s_comp} for matter with $T$ = 1 MeV
and $Y_P$ = 0.4.  More importantly, the composition of nuclear
matter, especially the wide spreads in distribution of nuclei,
could influence the position of the neutrinosphere and the spectra of radiated 
neutrinos and antineutrinos.

The complete equation of state table is not only useful for core
collapse supernova simulations, but also for other astrophysical applications, including the long term evolution of proto-neutron stars, and neutron star mergers.  Our procedure, for matching mean field and virial calculations to produce a thermodynamically consistent EOS, can be used for a variety of effective interactions.  In this paper we have presented an EOS based on the NL3 interaction that is relatively stiff at high densities.  In future work we will present additional EOS tables based on softer interactions that have lower pressures at high densities. 

Finally, our full EOS tables, both with and without lepton and photon contributions, are available for download as described in Appendix \ref{app}.

%\section{Acknowledgement}

We thank Lorenz H$\ddot{\mathrm{u}}$edepohl, Thomas Janka, Andreas Marek, Evan O'Connor, and Christian Ott for important help running astrophysical simulations to debug our equation of state. This work was
supported in part by DOE grant DE-FG02-87ER40365. 

\section{\label{app}Appendix: format of EOS tables.}

Here we describe the physical quantities provided in the tables
for the equation of state, NL3eos1.03.dat and NL3eosb1.03.dat and where they can be downloaded.   One should download the gzip compressed files NL3eos1.03.dat.gz and or NL3eosb1.03.dat.gz (that are about 100 MB each) and use gunzip to decompress them.  The grid structures of these tables are indicated in Table \ref{tab:phasespace2} and contain approximately 517 MB of data each.     The tables, a sample FORTRAN computer program, and a readme file are available for download both at our website 
\begin{widetext}
\underline{\url{http://cecelia.physics.indiana.edu/gang_shen_eos/}} 
\smallskip
\end{widetext}
and, from the  Electronic Physics Auxiliary Publication Service - EPAPS web site, \cite{EPAPS}.   Please check our web site for any updated information regarding these EOS tables.

The tables contain a number of quantities.  For a single triplet of $T$, $n_B$, and $Y_P$, there are 16 items that make up a row of the table.  These items are described below.  In the table NL3eosb1.03.dat, only the contribution from baryons is taken into account for items 4,5,6. In the table NL3eos1.03.dat, the contributions from electrons, positrons, and photons are also included. The electron mass is 0.511 MeV.  

\subsection{Items in EOS tables.}
\begin{enumerate}
\item Temperature $T$ [MeV]. The range of
temperature is first $T$=0 then from $10^{-0.8}-10^{1.875}$ MeV, see Table \ref{tab:phasespace2}.     \item Proton fraction $Y_P$. The range of proton fraction is first 0, and then from, 0.05 $\sim$ 0.56. The step in proton fraction is 0.01, see Table \ref{tab:phasespace2}.
    \item Baryon number density $n$ [fm$^{-3}$]. The range of
    density is from 10$^{-8}$ to 10$^{0.175}$ fm$^{-3}$. See Table \ref{tab:phasespace2}.
    \item Free energy per baryon $F/A$ [MeV] which has subtracted the
    free nucleon mass 939 MeV.
    \item Pressure $P$ [MeV/fm$^{3}$].
    \item Entropy per baryon $S/A$ [k$_B$].
    \item Chemical potential for neutrons $\mu_n$ [MeV]. The tabulated value
    is relative to the nucleon mass 939 MeV.
    \item Chemical potential for protons $\mu_p$ [MeV]. The tabulated value
    is relative to the nucleon mass.
        \item Chemical potential for electrons $\mu_e$ [MeV]. The tabulated value 
    includes the electron mass.
    \item Average mass number $\bar{A}$ of heavy nuclei with $A>4$, which exclude
    alpha particles.
    \item Average proton number $\bar{Z}$ of heavy nuclei with $A>4$, which exclude
    alpha particles.
    \item Mass fraction of free neutrons.
    \item Mass fraction of free protons.
    \item Mass fraction of alpha particles.
    \item Mass fraction of heavy nuclei with $A>4$, which exclude
    alpha particles.
    \item Effective nucleon mass $M^*$ [MeV]. In uniform matter it
    is obtained from RMF theory. For virial gas and non-uniform
    matter, it is chosen to be the free nucleon mass $M^*=939$ MeV.
\end{enumerate}

\subsection{Sample FORTRAN computer program readeos.f}

The Fortran program readeos.f includes a very short main program that calls the subroutine load\_table, to read NL3eos1.03.dat or NL3eosb1.03.dat, and then calls the subroutine readeos with inputs $T$ (in MeV), proton fraction $Y_p$, and density $n$ (in fm$^{-3}$).  The subroutine readeos uses triliner interpolation (in $T$, $Y_p$, and $n$) to return the above 16 values plus the internal energy per baryon (in MeV) and the chemical potential for electron neutrinos in chemical equilibrium (in MeV).  Note that one needs to call load\_table only once and then one can call readeos many times.  For further details please see the comments in readeos.f.

\end{document}